\documentclass{cta-author}

{}
{}
{}

\usepackage{subfig} 
\usepackage{url} 

\usepackage{pdflscape}

\usepackage{textcomp}

\usepackage{booktabs}
\newcommand{\tabitem}{~~\llap{\textbullet}~~}
\newcolumntype{R}[1]{>{\raggedleft\let\newline\\\arraybackslash\hspace{0pt}}m{#1}}
\newcolumntype{L}[1]{>{\raggedright\let\newline\\\arraybackslash\hspace{0pt}}m{#1}}
\newcolumntype{C}[1]{>{\centering\let\newline\\\arraybackslash\hspace{0pt}}m{#1}}

\begin{document}

\title{A review of smartphones based indoor positioning: challenges and applications}

\author{\au{Khuong An Nguyen $^1$, }
\au{Zhiyuan Luo $^2$, }
\au{Guang Li $^3$}
\au{Chris Watkins $^2$ }
}

\address{\add{1}{School of Computing, Engineering and Maths, University of Brighton, Brighton BN2 4AT, United Kingdom}
\add{2}{Computer Science Department, Royal Holloway, University of London, Surrey TW20 0EX, United Kingdom}
\add{3}{Institute of Cyber-Systems and Control, Zhejiang University, Hangzhou 310027, China}
\email{K.A.Nguyen@bton.ac.uk}}

\begin{abstract}
The continual proliferation of mobile devices has encouraged much effort in using the smartphones for indoor positioning. This article is dedicated to review the most recent and interesting smartphones based indoor navigation systems, ranging from electromagnetic to inertia to visible light ones, with an emphasis on their unique challenges and potential real-world applications. A taxonomy of smartphones sensors will be introduced, which serves as the basis to categorise different positioning systems for reviewing. A set of criteria to be used for the evaluation purpose will be devised. For each sensor category, the most recent, interesting and practical systems will be examined, with detailed discussion on the open research questions for the academics, and the practicality for the potential clients.
\end{abstract}

\maketitle

\section{Introduction}
Indoor positioning is the technology that helps locating objects or guiding people in unfamiliar, complex buildings. Despite its helpfulness, it has been almost two decades since GPS was introduced for outdoor positioning, yet the search for an equivalent ubiquitous, affordable, and accurate indoor counterpart is still going on. The challenge for such technology stems from the complex indoor interior design, and the building materials which block and distort the radio, satellite signals.

For the past five years, the proliferation of the mobile devices, along with the continuing miniaturisation of sensors, have propelled the smartphone as an potential instrument for future indoor positioning systems. Mobile phones have now eclipsed desktop computers in terms of worldwide market share\footnote{https://gs.statcounter.com/platform-market-share/desktop-mobile/worldwide - last accessed in 5/2020}. For the developers, smartphones are not just mini-computers, but are also sensing devices with full awareness of its surroundings, providing a complete yet compact computing package. For the consumers, they need not buying new hardware, nor having to carry an extra device just for the positioning service. 

\begin{figure}[h]
    \centering
    \includegraphics[width=3.3in]{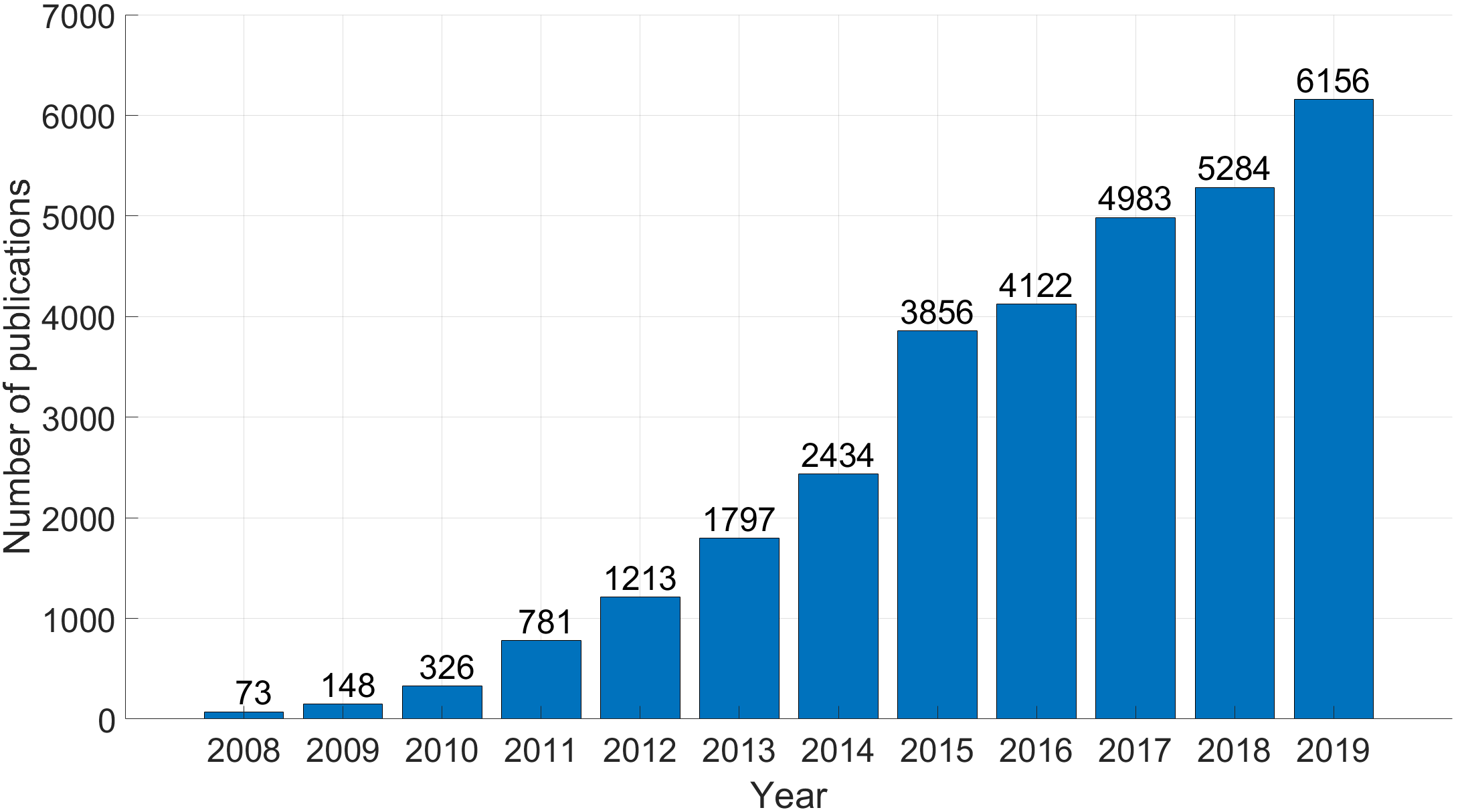}
    \caption{The estimation of the number of research work on smartphones indoor positioning indexed by Google Scholar. The search criteria was that the paper must contain both the `indoor' and `smartphones' keywords, and either `navigation' or `positioning' or `localisation' or `tracking', within the text. The results were then filtered by the authors based on their relevance.}
    \label{publications}
\end{figure}

Since 2005, the number of research papers involving smartphones indoor positioning has steadily increased (see Figure~\ref{publications}), which implies the overwhelming interest on this topic. At the time of writing, there have been over 30,000 related papers indexed by Google Scholar since late 2007 when the first iPhone and Android were released. Hence, to cater for the research community, there has been a wealth of surveys spreading across the research domain in the past decade. However, many of which are either too broad by covering most technologies in a general sense including impractical ones, or too narrow by focusing on a certain niche market using proprietary devices that most consumers do not possess or cannot afford. Therefore, this article is dedicated to the emerging smartphones based systems, which inherit a massive user base as well as the high level of interest from the academic researchers. At the end of the article, we aim to answer the following research questions.
\begin{itemize}
    \item \textbf{What is the most promising, practical smartphones based system for mass adoptions?} This review will judge the system's practicality on the same set of criteria.
    \item \textbf{What are the most interesting open questions for each system category?} Academics working on smartphones based indoor positioning may find interesting novel ideas for future research.
\end{itemize}

\subsection{Article's contributions}
With the growing interest in the domain, and the high number of relevant research outputs, it would be beneficial for the research community to have a review article every 3 to 5 years. In addition to this fact, our article offers the following contributions.
\begin{itemize}
    \item Only smartphones sensor based systems were reviewed (i.e. ultrasound, laser scanner, etc. which are not present on the smartphones, are omitted). All sensors are discussed from the smartphones' perspective, and how they may be adopted for the positioning purpose.
    \item We emphasise on the practicality of each technology, and the open research questions.
    \item We devise a taxonomy to group different smartphones sensors.
\end{itemize}

\subsection{Style and structure of this review}
To cater for the readers with interest in different systems, this article is written in a balanced narrative and systematic fashion, where a detailed, comprehensive literature survey is carried out, yet it is structured in a coherent, well formatted template, allowing the readers to pick up any chapter independently. The mathematics are kept to a minimum, with only well-proven formulas describing the foundation of the technique included.

For each sensor category, we describe the general underlying technology, followed by how it is used specifically on the smartphones, including the provided measures and the challenges. At the end of each category chapter, we review the positioning methods that are applicable to this class of sensor, from a theoretical viewpoint with open research questions targeted at interested academic researchers. Then, we compare the performances of those techniques implemented on real world systems as reported in the literature, aimed at practical implementations.

The remaining of this article is structured as follows. Section 2 introduces the concepts of smartphones based positioning. Section 3 devises a taxonomy of smartphones sensor categories along with the criteria to review them. So that, Sections 4, 5, 6, 7, 8 and 9 will review individual category in detail. Lastly, Section 10 summaries the article and outlines further work.

\begin{table*}[!ht]
\caption{The applications of smartphone indoor positioning system.}
\centering 
\begin{tabular}{ l L{7.5cm} L{7.5cm} } 
\toprule
 & \multicolumn{1}{c}{\textbf{Tracking purpose}} &  \multicolumn{1}{c}{\textbf{Navigation purpose}} \\ [0.5ex]
\midrule 
\textbf{Security} & For the guards:  & For the guards: \\
& \tabitem Being notified if the precious object has been moved. & \tabitem Finding the shortest way to the stolen/lost objects. \\
&  &  \\
& For the clients: & For the clients: \\
& \tabitem Automatically granting access for authorised personnel. & \tabitem Quickly evacuating personnel during an emergency.  \\ [0.5ex]
\midrule 
\textbf{Healthcare} & For the patients: & For the patients: \\
& \tabitem Locating the nearest wheel-chair. & \tabitem Route-finding to the doctor's office. \\
& \tabitem Automatic checking-in/out upon entering the hospital. & \tabitem Automatic wheel-chair navigation in the hospital. \\
&  & \\
& For the doctors:  & For the doctors: \\
& \tabitem Locating medical equipment. & \tabitem Finding the shortest route to the patient in an emergency. \\
& \tabitem 24 hour monitoring of mental illness patients. &  \\ [0.5ex]
\midrule  
\textbf{Retailer} & For the customers:  & For the customers: \\
& \tabitem Shopping assistant for the latest in-store offers. & \tabitem Finding the correct shelf for a particular item. \\
&  & \\
& For the managers: & For the managers: \\
& \tabitem Sending out relevant e-coupons when the customers are near a particular shelf. & \tabitem Improving store layout by analysing footfall and congestion. \\
& \tabitem Pinpointing staff's location in realtime. &  \\ [0.5ex]
\midrule  
\textbf{Industry} & For the customers:  & For the customers: \\
& \tabitem Tracking luggage at the airport. & \tabitem Finding seats in large venue.  \\
& & \tabitem Personalised computer guided tours. \\
&  & \\
& For the administrators:  &  For the administrators: \\
& \tabitem Locating products in the warehouse. &  \tabitem Automatic adjusting the speed of conveyor belt for transporting heavy items. \\
& \tabitem Be notified when an item leaves the warehouse. &  \\ [0.5ex]
\bottomrule 
\end{tabular}
\label{applications}
\end{table*}

\section{Smartphones based indoor positioning}
This section introduces the concept of smartphones indoor positioning, highlighting the applications and challenges facing such systems.

\subsection{Tracking with smartphones}
For this type of system, we assume that the user must carry a smartphone with them at all times, in order to deliver the positioning service. This assumption is justified, given that 3.5 billion people possessing a smartphone in 2019, and over 94\% of adults in the UK have one, according to a recent survey\footnote{https://www.statista.com/statistics/300378/mobile-phone-usage-in-the-uk - last accessed in 5/2020.}.

However, smartphones do not come ready with indoor tracking capability. Thus, most solutions rely on an app running in background to deliver such service. This app receives data from the built-in sensors, some of which require the user's permission at launch time. Some systems may require additional supporting infrastructure to interact with the phone app, as we will discuss in detail in the upcoming sections.

Compared to other indoor positioning competitors, smartphones based system offers unique benefits for the users in its compact form, and its self-contained package as a mini computing unit with an interactive touch screen. Most importantly, the clear advantage is being a ubiquitous device, the users need not carrying an extra piece of hardware for indoor positioning service.

\subsection{Challenges}
Despite the aforementioned advantages, it is worth emphasising that most smartphones sensors, apart from GPS, were originally designed for other functionalities in mind, rather than purposely for indoor positioning. Hence, we face the following challenges.
\begin{itemize}
    \item \textbf{The sensor design is minimal.} Over the years, sensors continue to be miniaturised in order to fit into the small phone body. The challenge is that their sensitivity drops consequently (i.e. a bigger sensor will have longer range with more accurate measures).
    \item \textbf{The measures are noisy.} This includes the mechanical noises as the sensors are packed closely together in tight space with other conductive components like battery, and unintended interference from external sources.
    \item \textbf{The sensors are heterogeneous.} The variety of smartphones designers, and sensor manufacturers, makes it challenging to normalise the measures across devices.
    \item \textbf{The interior design is complex.} The building material is made of large ferrous metal structures (e.g. metal bars, steel rebars, reinforced concrete) which greatly distort some sensor readings.
    \item \textbf{The indoor environment is dynamic.} The building is usually occupied by many users moving about in their daily work, which may impact some wireless signal based systems.
\end{itemize}

\subsection{Applications of smartphones indoor positioning system}
Overall, most indoor positioning systems have two general purposes, tracking and navigation. For the tracking purpose, at any moment's notice, the system must be able to detect and pinpoint the location of the person or object. For the navigation purpose, the system must guide the users using the most optimal route. Table~\ref{applications} compares some specific applications of indoor positioning for the four popular societal areas, which are healthcare, retailer, industry, and security.
\begin{table*}[!ht]
	\caption{Comparison of the characteristics of 18 most common smartphones sensors, in alphabetical order, for the indoor positioning purpose. The sensors are surveyed from the mid-range Lenovo Phab 2 phone.}
	\centering
	\begin{tabular}{ccccccC{6.5cm}}
		\toprule
		\textbf{Sensor type}	& \textbf{Temporal}	& \textbf{Spatial} &  \textbf{Battery} & \textbf{User} & \textbf{Max} & \textbf{Description}\\
		& \textbf{variation}	& \textbf{difference} &  \textbf{consumption} & \textbf{permission} & \textbf{frequency} & \\
		\midrule
		Accelerometer & low & high & low & none & 196 Hz & measuring the changing rate of the device acceleration \\ \addlinespace[0.2cm]
		
		Ambient light & low	& various & low & none & 4 Hz & measuring the ambient visible light's intensity \\ \addlinespace[0.2cm]

		Barometer & high & low & low & none & 90 Hz & measuring the atmospheric pressure \\ \addlinespace[0.2cm]		
		
		Bluetooth & high & low & low & yes & various & communicating with nearby Bluetooth beacons or other Bluetooth-enabled devices \\ \addlinespace[0.2cm]
		
		Camera * & various & high & high & yes & 60 Hz & capturing the scenery via the front or back lenses \\ \addlinespace[0.2cm]
		
		Cellular & low & high & high & yes & 20 Hz & communicating with nearby cell towers \\ \addlinespace[0.2cm]
		
		Fingerprint $^{\dagger}$ & low & N/A & low & yes & 0.5 Hz & generating an image of the finger's ridges and valleys \\ \addlinespace[0.2cm]
		
		FM $^{\dagger}$ & low & high & high & yes & N/A & receiving information from nearby radio towers \\ \addlinespace[0.2cm]
		
		GPS	& low & high & high & yes & 10 Hz & receiving the satellite signals to compute the latitude and longitude of the phone \\ \addlinespace[0.2cm]
		
		Gyroscope & low	& various & low & none & 198 Hz & measuring the changing rate of the device's tilting angle \\ \addlinespace[0.2cm]
		
		Heart rate $^{\dagger}$& N/A & N/A & low & yes & 1 Hz & measuring the pulse rate with reflected LED \\ \addlinespace[0.2cm]
		
		Magnetometer & low & various & low & none & 49 Hz & measuring the ambient magnetic field strength \\ \addlinespace[0.2cm]
		
		Microphone * & high & various & high & yes & 48 Hz & capturing the ambient acoustic noise \\ \addlinespace[0.2cm]
		
		NFC & various & various & high & yes & 1 Hz & communicating with nearby RFID tags \\ \addlinespace[0.2cm]
		
		
		Proximity & low	& various & low & none & 4 Hz & measuring distance to the nearest object within 10 centimetres \\ \addlinespace[0.2cm]
		
		Thermometer $^{\dagger}$ & low & low & low & none & 10 Hz & measuring internal phone components' temperature \\ \addlinespace[0.2cm]
		
		Time-of-flight $^{\dagger}$ & low & low & low & yes & 5 Hz & measuring distance to the nearest object within 2 to 3 metres\\ \addlinespace[0.2cm]
		
		WiFi & high	& low & high & yes & 0.03 Hz & communicating with nearby Access Points or other WiFi-enabled devices \\ \addlinespace[0.2cm]
		\bottomrule
	\end{tabular}
	
	* these sensors must operate in the foreground at all time. \\
	$^{\dagger}$ these sensors may not be available in all devices.
	\label{comparisonsensors}
\end{table*}

\section{Classification of smartphones sensors}
Having discussed the general idea of smartphones based positioning, we are now in a good position to delve deeper into the technological details of each individual sensor.

\begin{figure}[!ht]
	\centering
	
	\subfloat[The front of the phone.]{\includegraphics[height=2.3in]{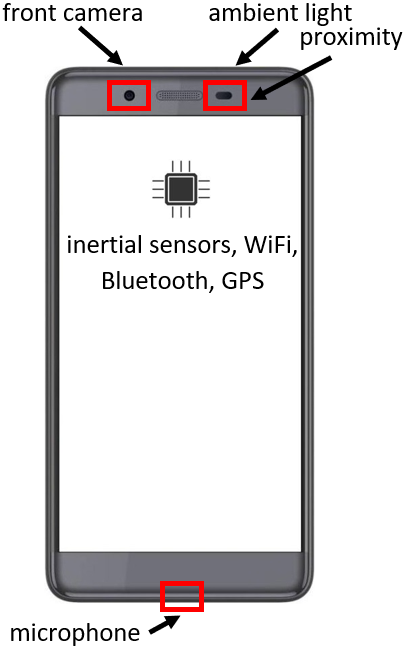}
		\label{frontphone}}
	\subfloat[The back of the phone.]{\includegraphics[height=2.1in]{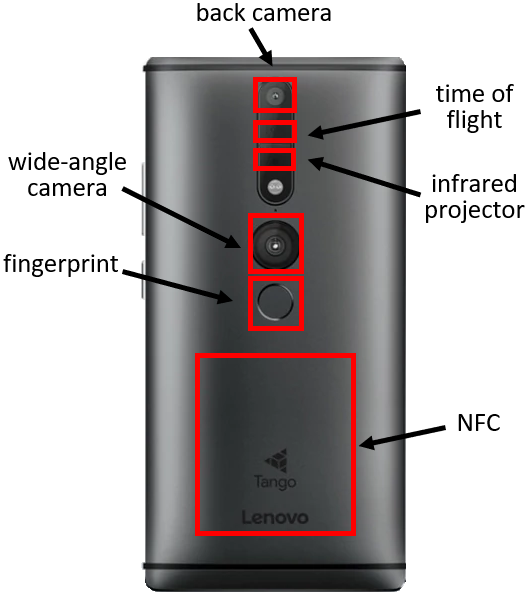}
		\label{backphone}}
	\hfil
	
	\caption{Relative position of most smartphones sensors, illustrated on our Lenovo Phab 2 test phone. Some sensors are embedded within the phone, while others are exposed. The battery occupies the majority of space at the bottom half of the phone.}
	\label{frontback}
\end{figure}

In short, a sensor is an electronic device that measures the changes in electrical or physical signals, and produces a measurable digital response to those changes. These changes can either be internal reactions within the mobile device, or externally in the surrounding environment. There are currently 18 sensors in modern smartphones (see Table~\ref{comparisonsensors} and Figure~\ref{frontback}). Some of which are more useful for indoor positioning than others. In particular, we identify five properties that are most relevant for our purpose.
\begin{itemize}
    \item \textbf{Temporal variation.} This is the sensor's ability to produce consistent measures over time, under the same environmental setting (e.g. in the same location).
    \item \textbf{Spatial difference.} For indoor positioning, it is critical that distinct locations have distinguishable sensor signatures, which are useful for pattern matching algorithms.
    \item \textbf{Battery consumption.} The lower the energy consumption, the longer the positioning app may continue running.
    \item \textbf{User permission.} Some sensors such as inertial ones can be queried at any moment, without a consent request, which enables smoother user experience. Others require special attention from the user in the form of a pop-up window to grant access (e.g. WiFi, Bluetooth, GPS).
    \item \textbf{Sampling rate.} This metric determines how frequent the sensor updates its measure. A high sampling rate sensor is critical for tracking fast moving users and objects.
\end{itemize}



\begin{figure*}[!ht]
\centering
\includegraphics[width=6.2in]{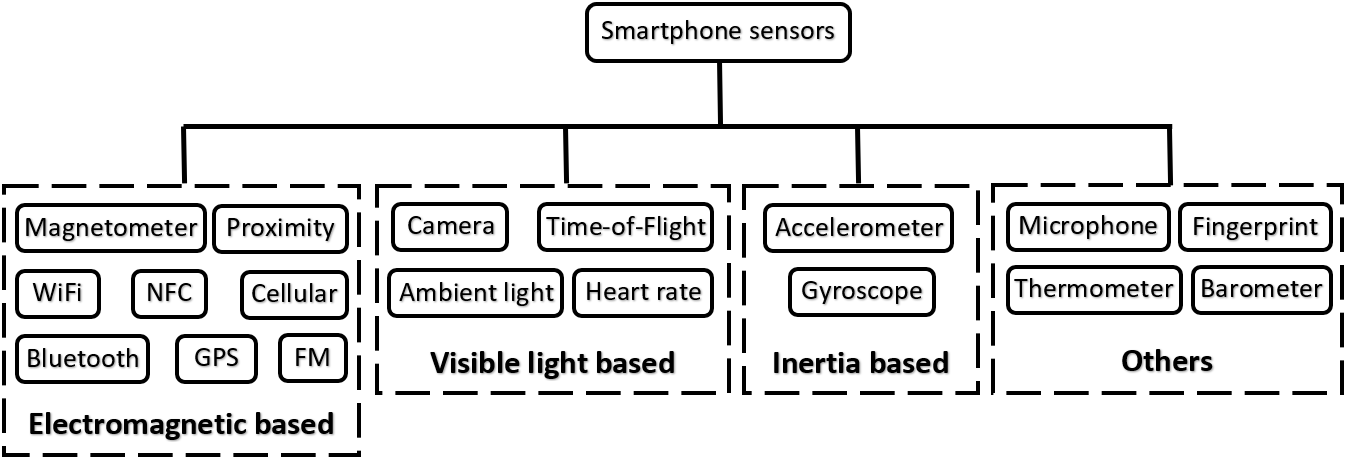}
\caption{The taxonomy of smartphones sensors.}
\label{smartphonesensors}
\end{figure*}

However, reviewing so many different sensors individually would make it inconvenient for the readers to follow, not least some of them sharing common properties, and many techniques may be applied to multiple sensors. Hence, we derive a taxonomy to categorise those sensors into four groups based on their functioning mechanism, which are electromagnetic based, visible light based, inertia based, and other sensors (see Figure~\ref{smartphonesensors}).
\begin{itemize}
    \item \textbf{Electromagnetic based}. These sensors operates on the electromagnetic spectrum, through the invisible waves of wireless signals.
    \item \textbf{Visible light based}. These sensors rely on visible natural lights to function.
    \item \textbf{Inertia based}. These sensors uses motions to estimate the phone's position.
    \item \textbf{Others}. This group includes the microphone, fingerprint, thermometer, and barometer, which do not directly fit in the above groups.
\end{itemize}

The sensors from each group will be reviewed together in their own separate section. In each section, we will overview the technologies behind each system category, from the smartphones' perspective, emphasising on their challenges, the positioning algorithms, and the results of the most notable systems in the recent literature.

\section{Electromagnetic signal based systems}
\begin{figure*}[h]
\centering
\includegraphics[width=5in]{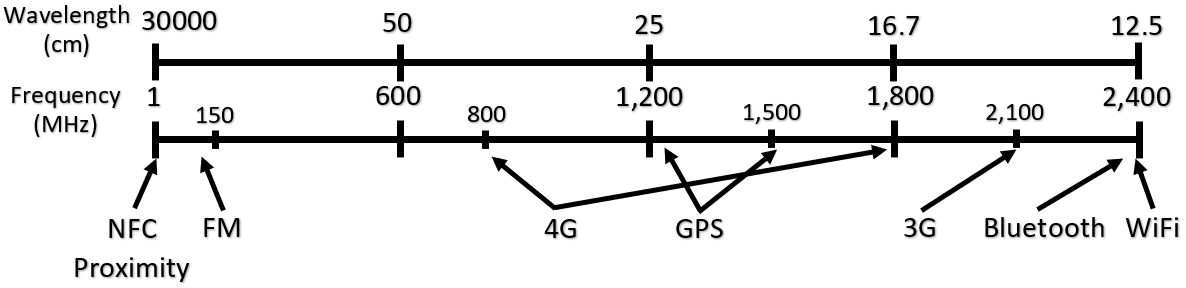}
\caption{The electromagnetic spectrum.}
\label{electrospectrum}
\end{figure*}

This section examines the systems that operate on the electromagnetic spectrum, with frequency ranging from as low as 10 Hz to the crowded 2.4 GHz band (see Figure~\ref{electrospectrum}). These electromagnetic waves carry information, generated by both electric and magnetic components, across the space without the need of any transport medium, and can be interpreted by the smartphones sensors. Some systems are ubiquitous as the transmitters already exist for other purposes, while others require dedicated hardware deployment, to be discussed later on.

Distance wise, the lower the frequency of the system is, the longer the wavelength is (i.e. the further the signal can travel, and the easier it can pass through walls, furniture). This attribute demonstrates why low frequencies are reserved for AM, FM, Cellular transmitters, which need to reach long distance users. In contrast, the higher the frequency of the system is, the smaller the wavelength is (i.e. the shorter distance they travel). To compensate for the lack of distance, these systems have higher bandwidth capacity to serve many users simultaneously, which explains why WiFi, Bluetooth transmitters are on this frequency.

Data wise, electromagnetic signals are capable of carrying position related information from the transmitters (e.g. the MAC address, the wireless channel, geo-coded data, etc.) which the receiving smartphone may infer to determine its location.







\subsection{WiFi}
WiFi describes the technology that allows the local devices to communicate wirelessly via an Access Point (AP) on the 2.4 GHz band, or the newer 5 GHz band. It was designed to replace the wired Ethernet cable for faster and more secured data transmission. Modern buildings are populated with plenty of APs for internet access and wireless communication amongst the users, which inherently benefits WiFi-based systems as no extra hardware is needed.

On the smartphones, WiFi was intended to allow the user to connect to nearby hotspots or routers for Internet access, or the recent WiFi Direct protocol that allows two smartphones to exchange data directly without an AP. Therefore, the level of information exposure is not as expansive as on a PC or laptop. In particular, from the smartphones' perspective, the visible information are just the name, MAC address, and the received signal strength (RSS) of nearby APs. To obtain these information, the phone initiates a scan, which listens on each of the 14 WiFi channel on the 2.4GHz band for a short period of time for any data periodically sent by nearby APs. This process is known as passive scanning. By default, WiFi APs broadcast data simultaneously on all channels every 100 ms. Thus, the entire scanning process takes about 1.5 seconds on most smartphones. Other devices such as laptop or PC may initiate a faster active scanning process, that floods the probe request frames on all WiFi channels, and waits for the probe responses from the APs. The process normally completes within 200 ms to 500 ms, which is significantly faster than passive scanning. Unfortunately, active scanning is not supported on Android devices to preserve battery and reduce overloading the WiFi channels.

Therefore, the challenges for WiFi based system are:
\begin{itemize}
    \item \textbf{Indoor areas with similar RSS.} Since the APs were primarily designed with data communication in mind, their deployed positions may not be optimal for indoor positioning.
    
    \item \textbf{Changing indoor interior.} Any rearrangement of furniture may alter the way the WiFi signals are deflective, absorbed, and scattered around the building.
    
    \item \textbf{The missing APs.} As these systems rely heavily on the WiFi APs, an AP taken off-line may impact the positioning result, or worse, requiring a new survey of the building.
    
    \item \textbf{The scanning delay.} As discussed above, passive scanning on Android devices may take up to 1.5 seconds, which may not be suitable for monitoring fast moving users.
    
    \item \textbf{WiFi cell breathing.} This mechanism is implemented in most new WiFi network to permit overloaded APs with too many connected users to offload their work to neighbourhood APs. The consequence is that the coverage zone, and APs RSS, which are critical for a positioning system, dynamically change depending on the number of active users.
\end{itemize}

Overall, despite some of its challenges, WiFi is still perhaps one of the most popular choices for indoor positioning right now, due its ubiquity. 

\subsection{Bluetooth}
Bluetooth describes the wireless technology that allows devices to communicate directly to each other, on the same 2.4 GHz band as WiFi. Although both Bluetooth and WiFi appear to facilitate the same wireless communication purpose amongst devices, their design intention are different. Firstly, Bluetooth's target audience are small peripherals (e.g. mouse, keyboard) with a few concurrent connections (typically between only 2 devices), while WiFi's target is delivering high speed connection amongst many larger devices (e.g. laptop, PC). Secondly, Bluetooth devices are easy to switch on/off and be ready connect to others, whereas WiFi APs require excessive configuration. Thirdly, Bluetooth's coverage is much shorter at 5-10 metres, comparing to typically 30-50 metres from WiFi's. Lastly, Bluetooth was designed for power efficiency and affordability (i.e. a single 1.5V cell battery may power a Bluetooth beacon for up to 9 months), whilst WiFi's aim was to maximise high speed data transmission.

On the smartphones, Bluetooth was intended to connect the phone to peripherals such as headphone, fitness tracker, and to exchange small pieces of information where speed is a non-issue with other Bluetooth-enabled devices. Hence, similar to WiFi, the level of exposure is rather limited, in the form of just the MAC address and the RSS of nearby Bluetooth devices, which are obtainable from a scan. However, the major difference to WiFi is the duration of the Bluetooth scan, which depends on three parameters - the scan window (i.e. the length of a single uninterrupted scan), the scan interval (i.e. the gap between scans) of the phone, and the advertising frequency of the peripheral device (i.e. how often the device announces its existence on the 3 Bluetooth advertising channels). Once setup, WiFi sticks on one channel through out its life, whereas Bluetooth frequently hops between its 3 channels. Ideally, it is preferred to have a long scanning window with short interval, and a high advertising frequency to maximise the discoverability of nearby Bluetooth devices. However, the consequence is the high power consumption for both the phone and the peripheral device. By default, the most optimal Bluetooth scanning profile on Android (i.e. the SCAN\_MODE\_LOW\_LATENCY) sets both the window and the interval to 4,096 ms. This means the phone will scan uninterruptedly for just above 4 seconds, with another 4 second break till the next cycle. Therefore, the Bluetooth peripherals should be configured to advertise at least once every 4 seconds to be discovered by the phone.

Nevertheless, the challenges for Bluetooth based systems are:
\begin{itemize}
    \item \textbf{Deployment of Bluetooth beacons.} Having to manually install and find the optimal placements for all beacons as well as maintaining these devices are challenging.
    
    \item \textbf{Signal instability.} Due to the implemented frequency hopping technique, the Bluetooth signals from a beacon often fluctuate even when observed from the same location.
    
    \item \textbf{High latency.} The 4 second interval between scan may pose a challenge for applications requiring constant location update.
\end{itemize}

Overall, with its relatively short range and low power consumption, Bluetooth is suitable for beacon based solution.

\subsection{GPS}
Global Positioning System (GPS) describes the constellation of 24 satellites circling the Earth in a precise orbit two times per day~\cite{hofmann2012global}. The GPS sensor on the smartphones acts as a receiver to measure the distance from itself to the satellites, based on the receiving signal, and applies trilateration to determine the location of the phone (to be discussed in the positioning method section) using the satellites as reference points. In contrast to the above WiFi and Bluetooth technology, which were designed for the indoor environments, GPS was designed primarily for outdoor use.

From the smartphones' perspective, the available information are the phone's current longitude, latitude, timestamp, and accuracy (i.e. the radius of the circle with the phone at the centre), which are precisely what most positioning systems need. However, GPS-based indoor positioning faces the following challenges.
\begin{itemize}
    \item \textbf{Extremely weak satellite signals.} Theoretically, the satellite orbits are arranged in such a way that at least four satellites are seen at any place on Earth. In reality, skyscrappers may block such view. Additionally, by the time the satellite signals reach the Earth's surface, they become too weak to penetrate most modern building materials.
    
    \item \textbf{Coarse positioning.} In normal working condition, GPS has an average of 5-10 metre accuracy, which is sufficient for outdoor navigation. However, such accuracy means the user may be estimated to be in a different room or building.
    
    \item \textbf{Excessive battery consumption.} GPS receiver on smartphones is known to drain battery quickly when used for long period of time.
\end{itemize}

Overall, in the indoor context, GPS is usually employed as opportunistic signal information, rather than a stand-alone indoor positioning solution.

\subsection{Cellular}
\label{cellular}
Cellular technology allows mobile devices to call, text each other, as well as accessing the internet via the mobile network provider. At the heart of this technology is the cell tower, which serves a specific geographical area (i.e. cell). When the users move between cells, their information are automatically handed over amongst towers. To avoid signal interference, adjacent cells use different signal frequencies.

From the smartphones' perspective, the available information are the cell ID, the signal strength, and the signal to noise ratio of nearby cell towers. Most notably, there are multiple large crowd-sourced cell tower databases\footnote{https://www.opencellid.org - last accessed in 5/2020}\footnote{https://radiocells.org - last accessed in 5/2020}, which contain the geographical location of most towers, to be used for trilateration of the mobile devices.

Nevertheless, cellular-based indoor positioning faces the following challenges.
\begin{itemize}
    \item \textbf{Highly coarse positioning.} It is common for cell towers to be placed 2 to 3 kilometres apart, making it challenging to distinguish the user location within a cell.
    
    \item \textbf{Weak signal.} Although cellular signals operate on lower frequencies than GPS, allowing it to penetrate most building materials to which the satellite signals struggle, well-shielded buildings with thick walls, glasses may still block out the cellular signals.
\end{itemize}

Overall, due to its low spatial signal difference, cellular technology may only offer city/town level positioning accuracy.

\subsection{FM}
\begin{figure}[h]
\centering
\includegraphics[width=3.3in]{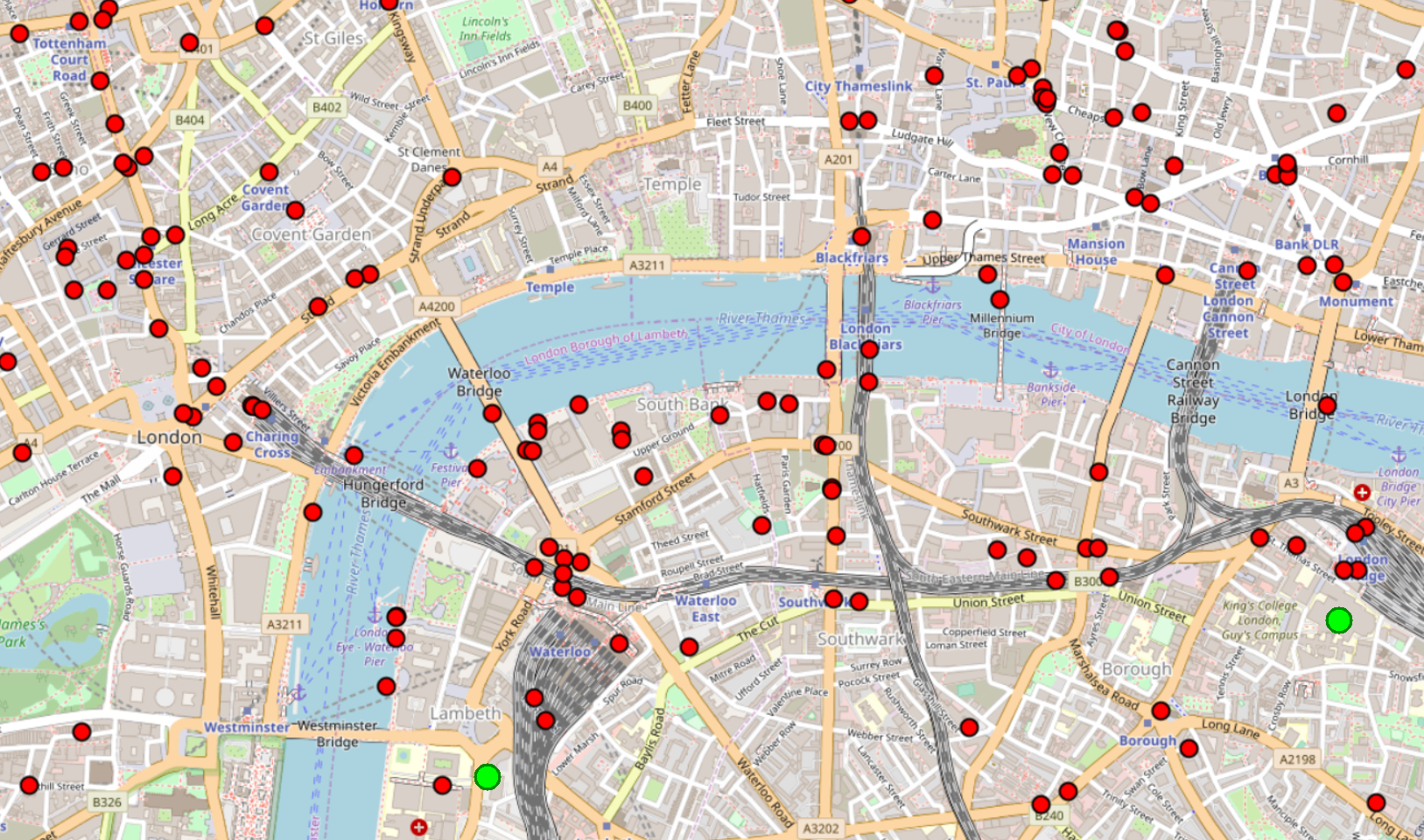}
\caption{The deployment of cellular towers of the UK O2 network and FM transmitters in central London. There are 141 cellular towers, depicted by the red dots, whereas there are only 2 FM transmitters, depicted by the green dots in the same area.}
\label{fmcellular}
\end{figure}

Frequency Modulation (FM), also widely known as FM radio, is the technology to broadcast sound-based information such as music, news to listeners with a receiver on the same frequency band. Each radio station occupies a fixed frequency on the FM spectrum, and has its own radio tower. Compared to cellular cell towers, FM radio towers are significantly more powerful transmitters, covering a much wider geographical area (see Figure~\ref{fmcellular}). 

The clear advantage of FM over other electromagnetic signals such as GPS, WiFi, Bluetooth, Cellular, is that it operates on significantly lower frequency, from 87.5 MHz to 108 MHz, making it easier to penetrate buildings and walls.

Nevertheless, since FM utilises the same concept of outdoor broadcasting tower as cellular, it suffers from the same weaknesses, which the readers may refer to Section~\ref{cellular}. The biggest challenge for this type of system is that Android currently does not officially support the technology, and most APIs are vendor-specific, making it challenging for mass market deployment.

\subsection{NFC}
Near-field communication (NFC) describes a set of close range wireless protocols, enabling two devices within 10 cm to communicate. The technology is based on inductive coupling on the 13.56 MHz frequency, between the antennas on NFC-enabled devices. 

From the smartphones' perspective, NFC is made popular by contactless payment applications (e.g. Apple Pay, Google Pay), which allow the consumers to make payment on the go by tapping their phones onto a payment device. The technology also allows smartphones to read information (e.g. room number, geo-coded position) embedded on NFC tags attached on walls, doors, furniture. The advantages of this approach are that NFC tag is passive with no internal power, and is rather inexpensive to deploy.

However, NFC based systems face the following challenges.
\begin{itemize}
    \item \textbf{Short communication range.} The active working distance is just within 10 cm. Hence, the users must almost touch their phone to an NFC tag or device, which impacts the hands-free feature of the system.
    
    \item \textbf{Limited simultaneous reading.} Android allows only one tag to be read at a time, although the technology is capable of observing up to 4 tags stacked on top of each other, simultaneously.
\end{itemize}

Overall, because of the short working distance within 10 cm, NFC based solutions may not be suitable for the indoor positioning purpose yet.

\subsection{Proximity}
The proximity sensor is designed to detect objects within its vicinity without any physical contacts. To do so, it emits an electromagnetic beam (e.g. infrared LED) and senses any light that bounces back. The more light being reflected back, the closer the object should be. The active working range is fairly short at about 6 cm, which is slightly shorter than NFC's.

On the smartphones, the proximity sensor is situated by the front camera and the ear speaker (see Figure~\ref{frontback}), with the sole purpose of automatically turning off the screen while the user is holding the phone by their ears (by detecting the side of the user's face) to avoid unintended touches. Although the sensor is capable of estimating the distance to the nearest object in centimetre, some smartphones' firmware restricts the output in simple binary form of being near or far.

At present, because of the inconvenient placement and the restricted output, this sensor does not appear to be useful for indoor positioning. Consequently, there has been no published literature that employs such technology. Nevertheless, we include it here for completeness.

\subsection{Magnetometer}
\label{magnetometer}
The magnetometer is designed to measure the strength and direction of the static magnetic fields (e.g. the Earth's magnetism, the magnet bar). These static fields do not change over time nor have a frequency (i.e. 0 Hz). Whereas, other electromagnetic signals discussed so far (e.g. WiFi, Bluetooth, GPS, FM and Cellular) are dynamic, in which their waves with different frequencies travel in the air. 

On the smartphones, this sensor was intended to identify the true North for outdoor navigation, made popular by the compass app. It may also be used as ferrous metal detector. Measurement-wise, the available information are the magnetic field strength, in microTesla unit $(\mu T)$, along the 3 axis of the phone. Modern device has a sampling rate of up to 100 Hz. The advantage of magnetic based system is that magnetism exists everywhere on Earth, and requires no additional infrastructure.

However, magnetometer based system faces the following challenges.
\begin{itemize}
    \item \textbf{Phone placement in 3D.} Since the sensor's frame aligns to the phone's body, any change to the phone's orientation in the 3-dimensional space varies the 3 sensor's measures, even in the same position.
    
    \item \textbf{Magnetic disturbance.} The compass does not work accurately indoors, because of the building materials (e.g. metal bars, steel rebars, reinforced concrete) which greatly distort the Earth's magnetic field.
\end{itemize}

Overall, magnetometer provides a means to capture the ubiquitous magnetic field that exists in all buildings, and may be valuable for some indoor positioning systems.

\subsection{Positioning algorithms}
Having understood the strengths and weaknesses of each electromagnetic based sensor, we are now in a good position to introduce the most popular positioning methods for these sensors, namely proximity tracking, trilateration, and location fingerprinting. At the end of the section, we will overview the most notable recent systems in the literature.

\subsubsection{Proximity tracking}
\label{electroproximity}
\begin{figure}[h]
\centering
\includegraphics[width=1.2in]{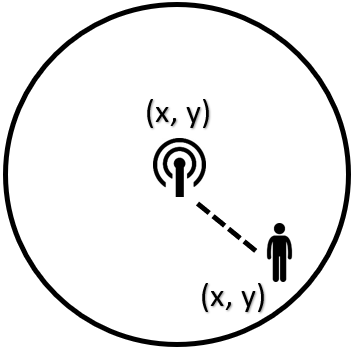}
\caption{Proximity tracking. The user location is assumed to be the position of the nearest (strongest) beacon.}
\label{proximity}
\end{figure}

This method aims at detecting whether the phone is close to a tracking beacon (e.g. the WiFi APs, Bluetooth beacons). It is suitable for systems in which only a binary decision about the presence of the tracking person in a certain area is needed (e.g. whether the customer is near a particular shelf to send out coupons, whether the user is inside a room to switch on the light), and not the precise fine-grained position. In these cases, the location of the person is assumed to be the known location of the tracking beacon (see Figure~\ref{proximity}). The sensors that benefit the most from this method is short range ones (e.g. Bluetooth), as they only broadcast signals within a small area.

The advantage of such approach is in its simple implementation, as there is no computation involved in. The disadvantage is clearly its limited positioning accuracy as the user is coarsely estimated within a wide area.

\subsubsection{Trilateration}
\label{electrotrilateration}
Trilateration improves on proximity tracking's accuracy by combining the RSS from multiple nearby beacons. It does so under the assumption that the position of at least 3 nearby tracking beacons is known. This is the technique that underlines modern GPS localisation~\cite{kolodziej2017local}.

\begin{figure}[h]
\centering
\includegraphics[width=2.0in]{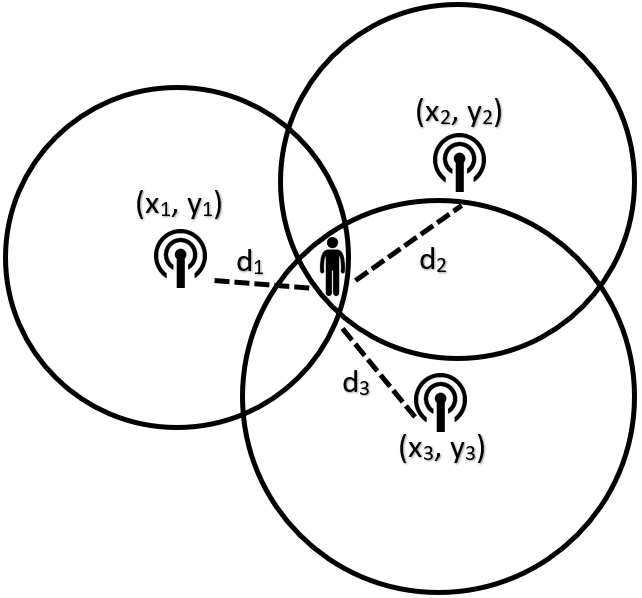}
\caption{Trilateration method. The user location is inferred as the intersection of the nearest beacons using geometry.}
\label{trilateration}
\end{figure}

It works in three steps. Firstly, the phone measures the RSS to the nearest 3 beacons. Secondly, for each beacon, it draws a circle representing the whereabouts of the phone with respect to the beacon (i.e. a weak signal will result in a small circle). Lastly, the intersection of these 3 circles is the estimation of the phone's position (see Figure~\ref{trilateration}). Without loss of generality, assuming the 2D co-ordinate of the three beacons are $(x_1, y_1), (x_2, y_2), (x_3, y_3)$, the distance $d_1, d_2, d_3$ between each beacon and the user is calculated based on the Pythagorean theorem as follows~\cite{pu2011indoor}:
\begin{equation}
    \begin{aligned}
        d_1 = \sqrt{(x_1 - x)^2 + (y_1 - y)^2} \\
        d_2 = \sqrt{(x_2 - x)^2 + (y_2 - y)^2} \\
        d_2 = \sqrt{(x_3 - x)^2 + (y_3 - y)^2} 
    \end{aligned}
\end{equation}

The intersected user position $(x, y)$ is then calculated as:
\begin{equation}
    \begin{aligned}
        x = \frac{A (y_3 - y_2) + B (y_1 - y_3) + C (y_2 - y_1)}{2(x_1 (y_3 - y_2) + x_2(y_1 - y_3) + x_3(y_2 - y_1))} \\
        y = \frac{A (x_3 - x_2) + B (x_1 - x_3) + C (x_2 - x_1)}{2(y_1 (x_3 - x_2) + y_2(x_1 - x_3) + y_3(x_2 - x_1))}
    \end{aligned}
\end{equation}
where $A = (x_1^2 + y_1^2 - d_1^2)$, $B = (x_2^2 + y_2^2 - d_2^2)$, $C = (x_3^2 + y_3^2 - d_3^2)$.

The distance between the user and each beacon $d_1, d_2, d_3$ can be obtained from the wireless signal strength (e.g. WiFi RSS) using several distance model, for example, the logarithmic distance path-loss model as follows~\cite{golestani2014improving,nowak2016path}:
\begin{equation}
    d_{1,2,3} = 10^{(\frac{TxPower - RSS} {10 * n})}
\end{equation}
where TxPower is the signal strength measured at 1 metre from a known beacon (e.g. -65 dBm), $n$ is the propagation constant (e.g. $n = 2$ for free space path-loss constant)~\cite{samimi2015probabilistic,sun2015synthesizing}, RSS is the measured signal strength from the user position, and $d_{1,2,3}$ is the distance between the beacon and the user (in metre).

The pros of trilateration is its practicality with simple geometry calculations. Its biggest cons, however, is the indoor signal attenuation. While such technique works great outdoors for GPS, with clear view to the satellites, there is often no line-of-sight from the phone to the indoor beacons. The signal is greatly attenuated that distorts the geometric shape of the broadcasting beacons which in turn does not provide accurate positioning estimation. A variation of trilateration is known as triangulation, which looks at the receiving signal's angle of arrival to compute the position. 

\subsubsection{Location fingerprinting}
\label{sec-fingerprinting}
Fingerprinting is arguably the most researched method in the indoor positioning literature. It addresses the weaknesses of proximity and trilateration methods by eliminating the need for both line-of-sight and the location of the beacons. The core concept of fingerprinting is based on a training database, generated to capture the wireless signal at every location in the building. Thus, the more unpredictable and challenging the signal propagation is, the more unique the training samples will be, which helps estimating the position of a user at an unknown location using his wireless signal~\cite{bahl2000radar}.

\begin{figure}[h]
\centering
\includegraphics[width=2.8in]{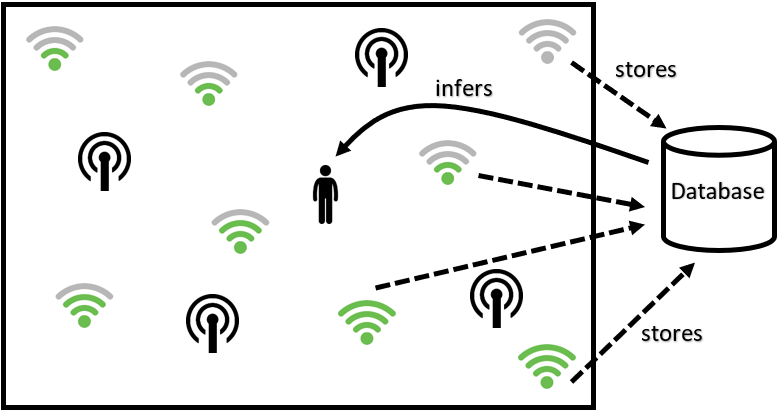}
\caption{Fingerprinting method. }
\label{fingerprinting}
\end{figure}

Fingerprinting has two phases, off-line surveying and online positioning estimation (see Figure~\ref{fingerprinting}). In the off-line phase, an expert surveys the building to generate a training database. He does so by walking through the building with a smartphone in hand to measure the wireless signal and the physical co-ordinate $(x,y,z)$ at every position. In the on-line phase, when the user wishes to discover his position, he measures the wireless signal at his current position. The system then searches through the training database generated previously to find a best match. This process is also known as pattern matching~\cite{bishop2006pattern,cormen2009introduction}.

\begin{figure}[h]
\centering
\includegraphics[width=3.3in]{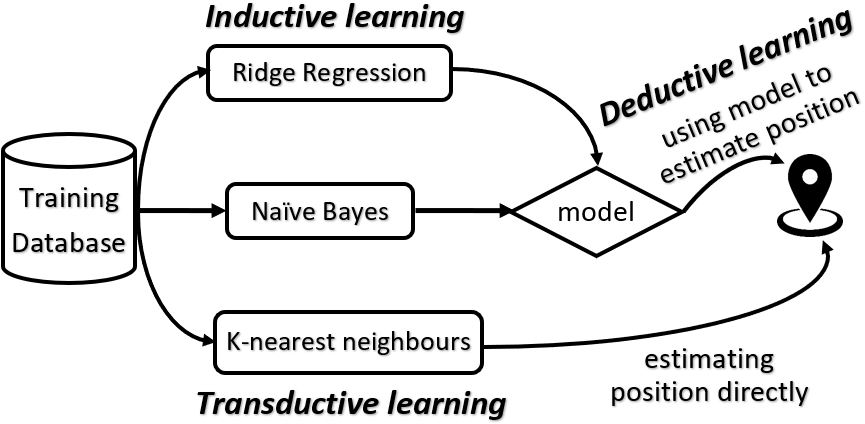}
\caption{The machine learning approaches for fingerprinting.}
\label{learningmodel}
\end{figure}

Fingerprinting is generally considered a supervised machine learning problem, as the training database contains both the object (i.e. the wireless signal strength) and the label (i.e. the Cartesian position). For this type of learning, there are two main approaches to estimate the positioning label. The first approach uses the training database to create a model, either probabilistic (e.g. Na\"ive Bayes)~\cite{madigan2005bayesian,zhao2019probabilistic} or deterministic (e.g. Ridge Regression, SVM)~\cite{nguyen2013enhanced,nguyen2015reliable}, which is then later used to estimate the position of a new signal sample~\cite{shalev2014understanding}. In contrast, the second approach skips the model learning phase to go directly from the training samples to estimating the position (e.g. K-nearest neighbours) (see Figure~\ref{learningmodel}).


Without loss of generality, the simplest form of fingerprinting can be formalised as follows. Given the training database with $M$ training examples $T_i=(\overrightarrow{RSS_i}, \overrightarrow{L_i})$ $(1\leq i\leq M)$, where $\overrightarrow{RSS_i}$ is the wireless signal strength, and $\overrightarrow{L_i} = (d_x^i, d_y^i, d_z^i)$ is the Cartesian positioning label of each training position $i$, the task is estimating the position $\overrightarrow{L_u}=(d_x^u, d_y^u, d_z^u)$ of a new wireless signal vector $\overrightarrow{RSS_u}$ somewhere in the building.

With the non-model based approach, the training example $T_i$ whose signal strength is nearest to the new sample's, is chosen as the estimated position. A popular measure metric for the RSS vector is the Euclidean distance:
\begin{equation}
\begin{aligned}
    \label{1.1}
    Eu(\overrightarrow{RSS_i}, \overrightarrow{RSS_u}) = Eu((AP^i_1, \dots, AP^i_N), (AP^u_1, \dots, AP^u_N)) \\ = \sqrt{\sum_{p=1}^N (AP^i_p - AP^u_p)^2}
\end{aligned}
\end{equation}
where $AP_i$ is the RSS measure of WiFi AP $i$, and $N$ is the number of APs.

However, as the WiFi signal attenuates, different positions may possess a similar RSS. Hence, the nearest training sample may not be the correct estimation. Thus, a set of $K$ nearest training samples are often considered. This approach is known as weighted K-NN, where a weight parameter is employed to penalise anomalous training samples as follows~\cite{honkavirta2009comparative,huang2011zigbee,shin2012enhanced}:
\begin{equation}
L_u = \frac{\displaystyle \sum_{i=1}^K \frac{1}{Dist(\overrightarrow{RSS_i}, \overrightarrow{RSS_u}) + \epsilon} \hspace*{2mm} \overrightarrow{L_i}}{\displaystyle \sum_{i=1}^K \frac{1}{Dist(\overrightarrow{RSS_i}, \overrightarrow{RSS_u}) + \epsilon}}
\end{equation}
where $\epsilon$ is a small positive constant to avoid division by zero.

With the model-based approach, we train a machine learning (ML) model using the provided training database. A popular algorithm is Ridge Regression which looks for a model that fits all training examples $T_i=(\overrightarrow{RSS_i}, \overrightarrow{L_i})$ $(1\leq i\leq M)$~\cite{draper2014applied}. This model will later be used to estimate the positioning label $\overrightarrow{L_u}$ for the new RSS vector $\overrightarrow{RSS_u}$. Ridge regression is based on mean squared error (MSE), with an extra tuning parameter called the ridge factor $\lambda \geq 0$ used to control the balance between fitting and avoiding the penalty. The ridge regression model is presented as follows:
\begin{equation}
\mathcal{M} = \sum_{i=1} ^ M {(\overrightarrow{L_i} - w \cdot \overrightarrow{RSS_i}) ^ 2} + \sum_{j=1} ^ N {\lambda ||w_j||^ 2}
\end{equation}
where $w_j$ $(1\leq j\leq N)$ are the weights for the $N$ WiFi APs. The closer $\lambda$ is to zero, the more linear the regression model is. In contrast, the bigger $\lambda$ is, the more weights close to zero or equal zero are. The objective is to find the optimal weights $w^*$ so that $\mathcal{M} \rightarrow \mathtt{min}$. The estimated position $\overrightarrow{L_u} = (d^u_x, d^u_y)$ for the new sample $\overrightarrow{RSS_u}$ is computed as follows, where each Cartesian dimension has a separated model
\begin{equation}
    \begin{aligned}
        d^u_x =  w_x \cdot \overrightarrow{RSS_u} \\
        d^u_y =  w_y \cdot \overrightarrow{RSS_u} \\
        d^u_z =  w_z \cdot \overrightarrow{RSS_u}
    \end{aligned}
\end{equation}

Another popular take to the model-based approach is modelling the training samples using a probabilistic model, in which the WiFi RSS are captured repeatedly at each training position to record the full signal distribution of each AP~\cite{bisio2016smart,murphy2012machine}. This model addresses the challenge of the WiFi RSS variation in the same position, causing a mismatch between the training RSS and the real-time RSS, despite being collected in the same place. Without loss of generality, given the new RSS sample $\overrightarrow{RSS_u}$, the algorithm calculates the posterior probability $P(T_i | \overrightarrow{RSS_u})$ of this RSS sample being observed at the training position $T_i$ $(1\leq i\leq M)$. The training sample with $\underset{i}{\max}(P(T_i | \overrightarrow{RSS_u}))$ is the estimated position. This probability is computed using Bayes' theorem~\cite{youssef2005horus}:
\begin{equation}
P(T_i | \overrightarrow{RSS_u}) = \frac{P(\overrightarrow{RSS_u} | T_i) \hspace{1mm} P(T_i)}{P(\overrightarrow{RSS_u})}
\end{equation}

Since $P(T_i)$ and $P(\overrightarrow{RSS_u})$ are given, all left is calculating the reverse probability $P(\overrightarrow{RSS_u} | T_i)$. By assuming the WiFi RSS from each AP are independent measures, we may apply the Na\"ive Bayes approach to calculate the posterior probability for each RSS as follows:
\begin{equation}
P(\overrightarrow{RSS_u} | T_i) = \prod_{j=1}^{N} P(AP_j^u | T_i)
\end{equation}
where $P(AP_j^u | T_i)$ presents how often the individual RSS of $AP_j$ appears at the training location $T_i$
\begin{equation}
P(AP_j^u | T_i) = \frac{how\hspace{1mm}many\hspace{1mm}times\hspace{1mm}AP_j\hspace{1mm}appears\hspace{1mm}at\hspace{1mm}T_i}{total\hspace{1mm}number\hspace{1mm}of\hspace{1mm}readings\hspace{1mm}observed\hspace{1mm} at\hspace{1mm}T_i}
\end{equation}

The major pros of fingerprinting is it exploits the complex indoor environment as its advantage. The more unpredictable the signal attenuation is, the more unique the training data will be, which aids the positioning estimation process. However, the major drawback of fingerprinting is the sheer amount of building space to be meticulously surveyed, the time it takes to perform such process, and the lack of physical reference for the training positions~\cite{nguyen2017performance}. As fingerprinting relies on a training database, below are some open research questions.
\begin{itemize}
    \item \textbf{How to alleviate the surveying process?} This is the most challenging procedure of fingerprinting, especially for large building.

    \item \textbf{How fine-grained the training database should be?} The building is normally segregated into evenly spaced grid where the calibration points are collected. The denser the points are, the bigger the training database will be, and the more efforts need to be invested.
    
    \item \textbf{How to obtain the position training label?} Each signal data needs a positioning label (e.g. the Earth's latitude and longitude, the Cartesian co-ordinate), yet it is not trivial to obtain such data. Most systems rely on an external system for ground-truth positions. Others assign them manually.
    
    \item \textbf{How to cope with indoor interior changes?} As the fingerprinting database is a representation of the signal propagation in the building at the time of measure, any interior changes (e.g. furniture rearrangement) may require a complete re-surveying of the building.
\end{itemize}


\subsection{Performance review}
\label{performanceelectromagnet}
Having discussed the most popular techniques in the last section, we are now in a good position to review their performances on real world systems. It is worth emphasising that this section only reviews single sensor based systems to expose the true performance related measure of each sensor (see Table~\ref{comparisonmagnet}). Hybrid systems will be reviewed later on in Section 9.

Technique wise, the overall consensus is that fingerprinting consistently achieved higher positioning accuracy at an average of 2-3 metres~\cite{zhao2019probabilistic,mok2007location,shin2012enhanced}, compared to others. Proximity and trilateration based systems, while being simple to deploy, could barely manage room-level accuracy at 5-6 metre accuracy~\cite{mok2007location,canton2017bluetooth}. Nevertheless, despite its high accuracy at the time of testing, fingerprinting based system may struggle to sustain such performance over long period of time, because any environmental changes may require the entire training database to be re-calibrated, as discussed earlier. Additionally, the density of the training points as well as the number of beacons also contribute to the positioning accuracy~\cite{kaemarungsi2012analysis,machaj2010impact,milioris2010empirical,chen2013placement,sharma2010access}.

Sensor wise, Bluetooth offers the highest accuracy, at 1.3 metres and 1.8 metres, 90\% of the time~\cite{nguyen2011robot,faragher2015location}. It is worth noting that these works strategically populated their own Bluetooth beacons densely for the training database. This observation was echoed by another work with the FM signal, where WiFi fingerprinting with more beacons, achieved double the positioning accuracy of FM fingerprinting, under the same testbed~\cite{moghtadaiee2011indoor}.

Overall, when it comes to system performance, WiFi and Bluetooth seem to be the two most popular options amongst electromagnetic sensors. WiFi, with its ubiquitous coverage in most modern buildings, offers slightly above room-level accuracy, whereas Bluetooth with its compact size, low power consumption, may offer 1-2 metre accuracy, when manually deployment is possible.
\begin{table*}[!ht]
	\caption{Comparison of notable electromagnetic signal based system performance.}
	\centering
	\begin{tabular}{C{2cm}cC{1.9cm}ccC{5.7cm}}
		\toprule
		\textbf{Authors} & \textbf{Sensor} & \textbf{Positioning technique} & \textbf{Testbed} & \textbf{Accuracy} & \textbf{Notes} \\
		\midrule
		
		Zhao et al.~\cite{zhao2019probabilistic} & WiFi & Probabilistic fingerprinting & 60m x 20m & 3.2m, 90\% chance & The authors slightly improved the positioning accuracy with filtering and AP selection. \\ \addlinespace[0.2cm]
		
		Shin et al.~\cite{shin2012enhanced} & WiFi & W-KNN fingerprinting & 48m x 22m &  5.3m, 90\% chance & Almost double the positioning accuracy compared to standard K-NN was reported. \\ \addlinespace[0.2cm]
		
		Mok et al.~\cite{mok2007location} & WiFi & Trilateration & 20m x 20m &  4m, 95\% chance & Only 5 WiFi APs were deployed in a relatively small area. \\ \addlinespace[0.2cm]
		
		Faragher et al.~\cite{faragher2015location} & Bluetooth & Fingerprinting & 600 $m^2$ &  1.8m, 90\% chance & Dense deployment of 1 beacon every 30 $m^2$. \\ \addlinespace[0.2cm]
		
		Nguyen~\cite{nguyen2011robot} & Bluetooth & Fingerprinting & 5m x 2.5m &  1.3m, 90\% chance & Dense deployment of 5 beacons in a small area. A robot was used to automate the surveying process. \\ \addlinespace[0.2cm]
		
		Canton et al.~\cite{canton2017bluetooth} & Bluetooth & Trilateration & 16.5m x 17.6m & 8.2m, 90\% chance & By exploiting the BLE channel diversity which is not available on the smartphones, the accuracy may be improved to 7m, 90\% chance. \\ \addlinespace[0.2cm]
		
		Moghtadaiee et al.~\cite{moghtadaiee2011indoor} & FM & W-KNN fingerprinting & 11m x 23m &  5.2m, 90\% chance & WiFi fingerprinting under the same testbed was reported with double the positioning accuracy. \\ \addlinespace[0.2cm]
		
		Viel et al.~\cite{viel2019original} & GSM & Fingerprinting & Udine city centre &  45m, 90\% chance & Different techniques including training models (Random Forest, SVM), KNN achieved similar positioning accuracy. \\ \addlinespace[0.2cm]
		
		Vandermeulen et al.~\cite{vandermeulen2013indoor} & Magnetometer & W-KNN fingerprinting & 14m x 16m &  6m, 90\% chance & If the room is known, the authors claimed 1.5m, 90\% chance. \\ \addlinespace[0.2cm]
		
		\bottomrule
	\end{tabular}
	\label{comparisonmagnet}
\end{table*}

\subsection{Overall remarks and open research questions}
Electromagnetic based approaches are perhaps the most researched category for indoor positioning. Its strength lies in the flexibility in catering for both infrastructure based and infrastructure free systems. As most modern buildings are already populated with many WiFi APs, the developers can utilise them, along with other opportunistic signals such as Celullar, FM, GPS, and the Earth's magnetic field~\cite{nguyen2015feasibility}. For developers preferring to deploy their own tracking beacons, Bluetooth APs offer an affordable solution with low power, and minimal beacon design.

Positioning accuracy wise, systems in this category normally achieve around room-level accuracy on average. Maintenance wise, depending on the chosen technique, there may be overhead in updating the training database, servicing the tracking beacons. 

Research wise, some interesting open questions for the academic researchers are:
\begin{itemize}
    \item \textbf{Electromagnetic interference.} The 2.4 GHz band in which WiFi and Bluetooth operate, is overcrowded with devices (e.g. phones, laptops, PCs) and home appliances (e.g. microwave, cordless phone, baby monitor). Collision in such environment is inevitable, which results in loss of connection, or beacons not visible during a scan.
    
    \item \textbf{Signal attenuation.} There is rarely a clear line-of-sight between the phone and the beacons. Hence, the wireless signals propagate unpredictably in the air. When two in-phase waves of signal meet, constructive interference results in a stronger signal, whereas, a combination of two out-phase waves will result in destructive interference which weakens both~\cite{nguyen2017performance}. The end receiving signal at the phone is a product of these phenomena. Besides, the user body which contains much water, is also a big attenuation factor~\cite{nguyen2011robot,nguyen2013evaluation}.
    
    \item \textbf{Device heterogeneity.} Various antenna from different smartphones have different sensitivities which impact the receiving signal. This has a major impact on systems relying on RSS.
\end{itemize}

\section{Visible light based systems}
This section discusses the systems that rely on natural, visible light to the human eyes, in contrast to the electromagnetic signal systems discussed in the last section, which use invisible waves of signal.

Distance wise, all systems in this category need a clear unobstructed sight from the sensor to the tracking person or object, as visible light cannot penetrate walls and furniture. Perhaps, this is the biggest challenge for this type of system.

Data wise, visible light (VL) does not physically carry information the way electromagnetic wave does. Light based systems rely on passively analysing the unique features of the scenery (i.e. whether there are enough distinct optical information amongst places), or actively illuminating the scene with their own light source and capturing the reflective light back (i.e. to measure the distance to the nearest object). Some systems used custom LEDs that flicker rapidly in unique pattern to distinguish amongst themselves, which can be decoded by the smartphones camera, yet are indiscernible to the human eyes. By altering the flickering frequency, information (e.g. the co-ordinate of the beacon) may be encoded within. This technique is known as Visible Light Communication (VLC).

\subsection{Camera}
Smartphones camera is CMOS imaging sensor, which is designed for low power device to capture an image of the surrounding. It does so by converting light waves passing through the lenses into digital signals. 

On the smartphones, camera was designed for either capturing scenery using the back facing camera, or taking selfie using the front facing camera. Hence, from the smartphones' perspective, the output is in the form of an image or video (i.e. a sequence of images). Modern smartphones are capable of recording up to 60 frames per second. The only assumption for camera based system is that there must be sufficient ambient light for the sensor to generate a clear image reflecting the surroundings.

Nevertheless, camera based systems face the following challenges.
\begin{itemize}
    \item \textbf{Phone placements.} Since the camera is fixed at the back of the phone, different angles will produce various perspectives of the same scenery.
    
    \item \textbf{Lighting condition.} The image sensor easily becomes under- or over- exposed, under different lighting.
    
    \item \textbf{No distance information.} Although the camera is capable of observing the world around, the captured information contain no depth information (i.e. it has no idea how far away an object is).
\end{itemize}

Overall, smartphones camera offers a means to see the world the way human do. It is currently one of the most popular choices for optical based systems.

\subsection{Time-of-flight}
Time-of-flight (ToF) is a 3D image sensor used to measure the distance to nearby objects. It tackles one of the biggest challenges facing standard CMOS cameras described earlier in the sense that everything captured in the image appears as in a flat world without any depth perception. ToF works by illuminating the scene with laser pulses, and observing the reflected light (i.e. the faster the light bounces back, the shorter the distance is)~\cite{nguyen2017assessing,li2014time}.

\begin{figure}[!ht]
	\centering
	
	\subfloat[Under normal lighting.]{\includegraphics[height=1.6in]{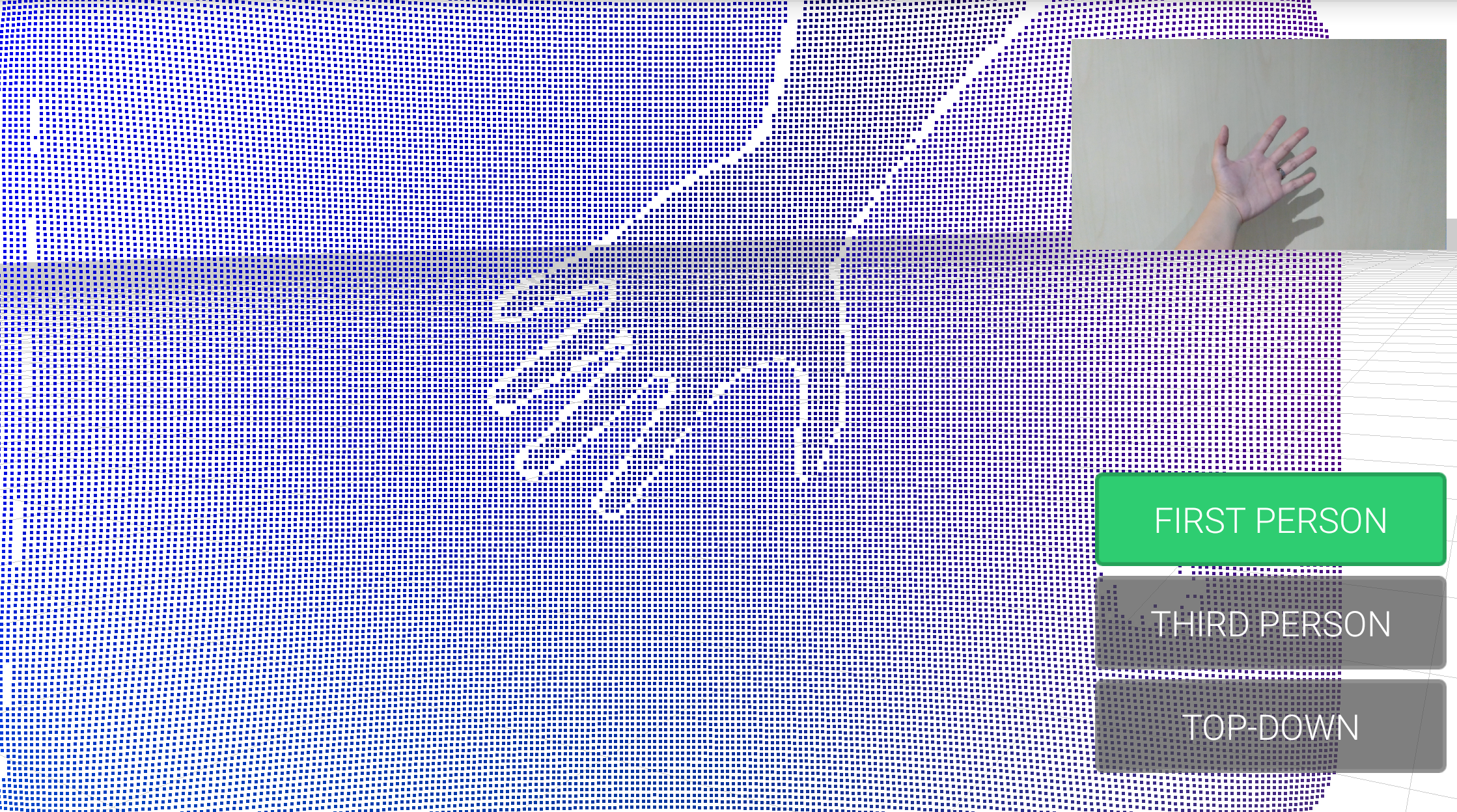}
		\label{tangobright}}
	
	\subfloat[Under dimmed lighting.]{\includegraphics[height=1.6in]{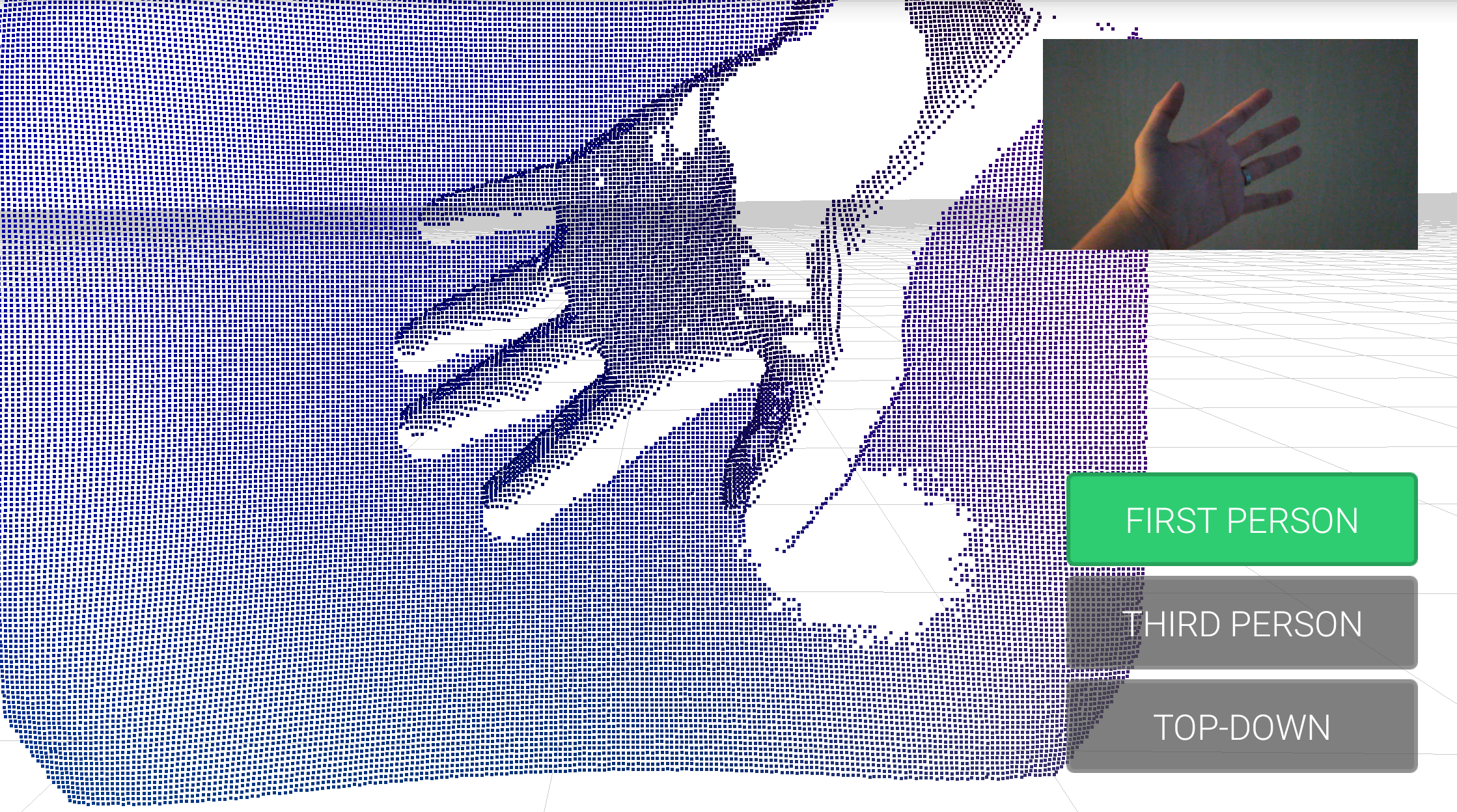}
		\label{tangodim}}
	\hfil
	
	\caption{The ToF point cloud image of a hand in different lighting conditions, observed by our Lenovo Phab 2 phone.}
	\label{tango}
\end{figure}

On the smartphones, ToF has two main functionalities, to scan the user's face for phone unlocking (using the front facing sensor), and to blur the background for portrait photography (using the back facing sensor). The former is much more popular in modern smartphones. Technically, the returned information is a point cloud containing the z-dimension depth information. These 3D points can be meshed together to create a surface rendering of the object and the environment (see Figure~\ref{tango}).

The big advantage of ToF is its high immunity to the ambient light, as it uses its own light source. However, the challenge for this type of technology is that only a handful of smartphones come equipped with this sensor. Additionally, ToF faces the following challenges.
\begin{itemize}
    \item \textbf{Sensor placement.} Most smartphones only employ the front facing ToF camera for facial recognition purpose. This set up is unlikely to be used by indoor navigation app to read the depth information of the scenery facing the other way, while the user is looking at the screen.
    
    \item \textbf{Reflective and absorbing surfaces.} As the distance measure is based on the reflected LED light, shiny surfaces (e.g. metal, glass) or absorbing surfaces (e.g. wood, plastic) make the measure inaccurate.
    
    \item \textbf{Slow frequency.} While standard CMOS camera can capture images at 60 frames per second or higher, current ToF cameras are restricted to 5 Hz, which may pose a problem for high speed scenes.
    
    \item \textbf{Limited range.} Most ToF sensors cannot see beyond 4 metres, as the LED beam becomes saturated beyond this range. 
\end{itemize}

Overall, a combination of ToF and standard camera would closely resemble the way human perceives the world around. As more smartphones come equipped with this sensor, it is a promising technology for future indoor positioning systems.

\subsection{Ambient light}
The ambient light sensor is a photodetector sensor, designed to detect the amount of light in the surroundings. It does so by converting the light energy into electric currents. On the smartphones, the sensor is exclusively used to adjust the screen's brightness with respect to the current ambient light.

The challenge for this sensor is the information it measures change significantly by various factor (e.g. the shadow of the human body, the facing angle of the sensor).

\subsection{Heart rate}
The heart rate sensor is an optical LED sensor, designed to measure the number of heartbeat per minute. As the sensor shines the LED light through the human skin, the blood pulses vary the light reflections. Hence, by measuring the amount of reflecting light, the number of heartbeats can be deducted. 

The challenge is that not all smartphones have this type of sensor. And at present, it does not seem to be useful at all for indoor positioning. Nevertheless, we include it here for completeness.

\subsection{Positioning methods}
Having discussed the strengths and weaknesses of all visible light based sensors, we will now overview the most popular positioning methods, and the most notable recent systems in the literature.

\subsubsection{Proximity and trilateration}
These methods are identical to the ones discussed for the electromagnetic based systems. Hence, we refer the readers to Sections~\ref{electroproximity} and~\ref{electrotrilateration}. In short, with proximity based tracking, when a person enters a sensor coverage area (e.g. a CCTV enabled zone), her presence will be recorded. With trilateration, as each sensor's coverage range is known, and the final position is deducted as the intersection of these areas.

\subsection{Triangulation}
\begin{figure}[h]
    \centering
    \includegraphics[width=1.8in]{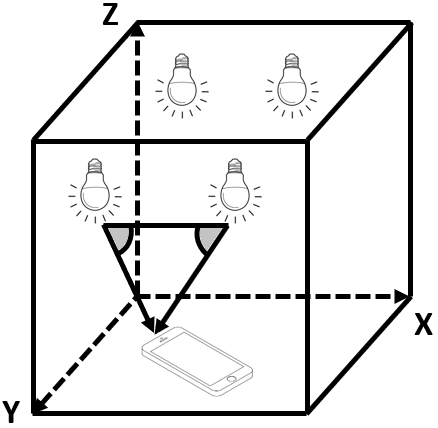}
    \caption{The triangulation approach with VLC.}
    \label{triangulation}
\end{figure}

The triangulation technique determines the location of the user by computing the angles in relation to at least two known fixed position beacons. In our case, the beacons are custom LEDs with unique flickering patterns, which are captured and decoded by the smartphones' front facing camera to work out its current position (see Figure~\ref{triangulation}). The working principle is inspired by the way mariners navigate on the sea, by observing the brightness from a set of fixed position stars in a constellation.

Compared to trilateration, which simply uses a rough estimation of the distance (i.e. WiFi RSS), triangulation is more knowledge-demanding as it requires not just the beacons position but also their spatial rotation. Additionally, because of the limited field-of-view of the smartphones camera, a dense LED grid needs to be deployed to guarantee visible beacons at any holding angle and position of the phone.

\subsubsection{Fingerprinting}
The concept of fingerprinting for visible light based systems are identical to that for electromagnetic based systems (see Section~\ref{sec-fingerprinting}). Instead of using the wireless RSS, optical information are used to create the training database. As the smartphones only provide image captures as outputs for each position, computer vision techniques were employed to extract notable features from such images. Ideally, these features must uniquely represent the position (i.e. no two positions should have the same set of features), as well as being robust to different lighting conditions. The most widely used technique is `Scale Invariant Feature Transform' (SIFT)~\cite{lowe1999object,bay2008speeded}.

The crux of this technique is finding locations (i.e. key points) within an image space that are invariant to image scaling or rotation, as well as being least affected by optical noises and distortions. 

Without loss of generality, given the image $I(x, y)$, the first step is defining a continuous function of scale known as scale space $L(x, y, \sigma)$ that is produced from the convolution of a variable-scale Gaussian $G(x, y, \sigma)$ as follows, with $*$ is the convolution operation and $\sigma$ is the variance of the Gaussian filter~\cite{lowe2004distinctive}.
\begin{equation}
    \begin{aligned}
        L(x, y, \sigma) = G(x, y, \sigma) * I(x, y) \\
        G(x, y, \sigma) = \frac{1}{2\pi\sigma^2}e^{-(x^2 + y^2)/2\sigma^2}
    \end{aligned}
\end{equation}

To detect key points in this scale space, the difference of two nearby scales is computed as follows~\cite{lowe1999object}
\begin{equation}
    D(x, y, \sigma) = (G(x, y, k\sigma) - G(x, y, \sigma)) * I(x, y)
\end{equation}
where k is a constant factor to separate the convoluted images in scale space.

For each image, each point is compared to its 17 neighbours in the current image and the ones above and below in the scale space. A point will be selected as a key point, if it is either larger or smaller than all of its neighbours.


\subsubsection{Visual odometry}
This method attempts to estimate the position and orientation of a moving phone by analysing the sequence of images taken by the on-board smartphones camera. It does so under the assumptions that there must be significant overlaps between continuous images (i.e. two images taken several seconds apart on a fast moving user will not contain sufficient overlaps), and the scenery must contain distinguishable textures (i.e. a big empty room with plain walls does not have unique information from any viewing angle).

Visual odometry has four steps~\cite{nister2004visual,scaramuzza2011visual}. In the first step, the image is captured by the smartphone's camera, whose notable features are then extracted. In the second step, the features between two consecutive images are compared (i.e. using features matching) based on some similarity metrics. Another approach for this step is called feature tracking, that is only extracting features from the first image, then track them in the next image with some correlation techniques. In the third step, the camera motion is estimated by either a Kalman filter, or finding the geometric properties of the features that minimise the cost function of the two consecutive images. In the last step, a time window is applied to refine the walking trajectory of the last $N$ frames.

Without loss of generality, the problem can be formulated as follows. Given a time-series of $N$ images taken by the smartphone camera $I=(I_1,\dots,I_N)$, the two camera positions $T_k$ and $T_{k-1}$ taken at two consecutive time $k$ and $k-1$ are related by the rigid body transformation as follows~\cite{scaramuzza2011visual}
\begin{equation}
    T_{k,k-1} = 
    \begin{bmatrix}
        R_{k,k-1} & t_{k,k-1}\\
        0 & 1
    \end{bmatrix}
\end{equation}
where $R_{k,k-1}$ is the rotation matrix, $t_{k,k-1}$ is the translation vector.

Finally, the camera poses are denoted as $C=(C_1,\dots,C_N)$, where $C_i, 1\leq i \leq N$ is the transformation of the camera at time $i$ with respect to the initial position at 0. The current camera pose $C_N$ is calculated by merging all transformation $T_k, 1\leq k \leq N$.

\subsection{Performance review}
\label{performancelight}
Having introduced the most popular techniques for VL based systems, we are now in a good position to review their performances on real world systems. The general impression is that most VL based systems achieved significantly higher positioning accuracy at centimetre level, than the previous electromagnetic based systems at metre level (see Table~\ref{comparisonlight}). More details will be discussed as follows.

Sensor wise, systems that use custom LEDs managed to achieve very high positioning accuracy at 22 centimetres and 40 centimetres, 90\% chance respectively~\cite{yang2015wearables,li2014epsilon}. However, it is worth noting that these systems employ dedicated LEDs that flicker at unique patterns. Realistically, it is unlikely that most venues will replace their existing lighting infrastructure with these custom LEDs just to support the indoor positioning service. It is preferable to deploy new beacons which are less disruptive, or ideally making use of the existing infrastructure. On this note, ToF system managed to achieve similar positioning accuracy, at 15.2 centimetres mean error~\cite{fang2015real}, without having to deploy additional LEDs. Nevertheless, very few current smartphones support ToF sensors, although this trend may change in the near future.

Technique wise, triangulation and trilateration, despite being simple to implement, excelled in this sensor category with very high centimetre positioning accuracy. However, these systems relied on additional hardware, as discussed previously, which adds the installation cost, and the disruption of setting up such hardware into the tracking area. On the other hand, fingerprinting based VL systems, which maintains a training database of either the lighting characteristics of the environment~\cite{zhang2016litell}, or the image of static landmarks~\cite{xiao2018indoor}, managed to achieve just above metre positioning accuracy. These systems conveniently fit into any building without altering the existing infrastructure, despite having to sacrificing a bit of the positioning accuracy, as well as the surveying labour cost, and the maintenance overhead of such database.
\begin{table*}[!ht]
	\caption{Comparison of notable visible light based system performance.}
	\centering
	\begin{tabular}{C{2cm}C{2.5cm}C{2cm}C{2.5cm}C{2cm}C{5cm}}
		\toprule
		\textbf{Authors} & \textbf{Sensor} & \textbf{Positioning technique} & \textbf{Testbed} & \textbf{Accuracy} & \textbf{Notes} \\
		\midrule
		
		Yang et al.~\cite{yang2015wearables} & 8 custom LEDs \& Camera & Triangulation & 2.4m x 1.8m x 3m & 22cm, 90\% chance & The authors employ polarization-based modulation to enable a low pulse rate positioning sytem. \\ \addlinespace[0.2cm]
		
		Li et al.~\cite{li2014epsilon} & 5 custom LEDs \& Camera & Trilateration & 5m x 8m & 40cm, 90\% chance & Channel hopping was employed where each LED beacon transmits via the optical channel which was claimed to be ambient light interference free. \\ \addlinespace[0.2cm]
		
		Zhang et al.~\cite{zhang2016litell} & Unmodified fluorescent lights \& Camera & Fingerprinting & 9000 $m^2$, 119 lights & 25cm, 90\% chance & The authors use normal lights in the building. Each FL has an inherent characteristic frequency which can serve as a discriminate feature. They claimed 10cm, 90\% chance while the user stands still. \\ \addlinespace[0.2cm]
		
		Xiao et al.~\cite{xiao2018indoor} & Camera & Fingerprinting & 1,000$m^2$ & 1.5m mean error & The authors built a training database containing image of static landmarks i.e. doors, windows. Experiments were done in a large open space museum. \\ \addlinespace[0.2cm]
		
		Fang et al.~\cite{fang2015real} & Time-of-flight & Visual odometry &  64.2m x 21.2m x 3.9m & 15.2cm mean error & The authors also tested the system in challenging dim environment with similar accuracy. \\ \addlinespace[0.2cm]
		
		\bottomrule
	\end{tabular}
	
	\label{comparisonlight}
\end{table*}





\subsection{Overall remarks and open research questions}
Overall, VL based system closely resembles the way human navigates. As a promising indicator, it offers some of the highest positioning accuracy at the time of writing, up to centimetre level with custom made LEDs, whereas infrastructure-free systems using just the smartphones camera may still offer 1 to 2 metre accuracy. Nevertheless, there are still challenges from the sensor design (e.g. smartphones camera has slow frame rate, narrow field of view, etc.), the lack of depth information (i.e. most current smartphones are not equipped with ToF sensor). With more and more smartphones come equipped with ToF sensor, VL systems may play a major part in future mainstream indoor positioning service.

Research wise, potential academic researchers may be interested in the following open questions:
\begin{itemize}
    \item \textbf{Lighting conditions.} As the smartphones camera completely rely on natural visible light to operate, over-exposure (i.e. too much lights) or under-exposure (i.e. too little light) will impact the captured optical information.
    
    \item \textbf{Viewing angle.} Most smartphone cameras have narrow viewing angle. In addition, the user tends to tilt the phone downwards while viewing the screen, which decreases the angle even further.
\end{itemize}

\section{Inertial based systems}
While previously discussed electromagnetic and light based systems rely on external references (e.g. radio signals, light, etc.) to function, inertial based systems use the phone motion to estimate the change of position in relation to the starting point. Those sensors include the accelerometer measuring the acceleration, and the gyroscope measuring the rotation. In the smartphones, those sensors are coupled together in one inertial measurement unit (IMU), often with the magnetometer described in Section~\ref{magnetometer}. The rationale is that although magnetometer does not directly measure motions, its function in detecting the Earth's magnetic North (e.g. as a compass) may be used in conjunction with accelerometer and gyroscope to determine the absolute heading. Nevertheless, the use of magnetometer for inertial tracking in this context could be restricted by the strong anomalies of the magnetic field within the building~\cite{nguyen2019location}.

\begin{figure}[h]
\centering
\includegraphics[width=1.7in]{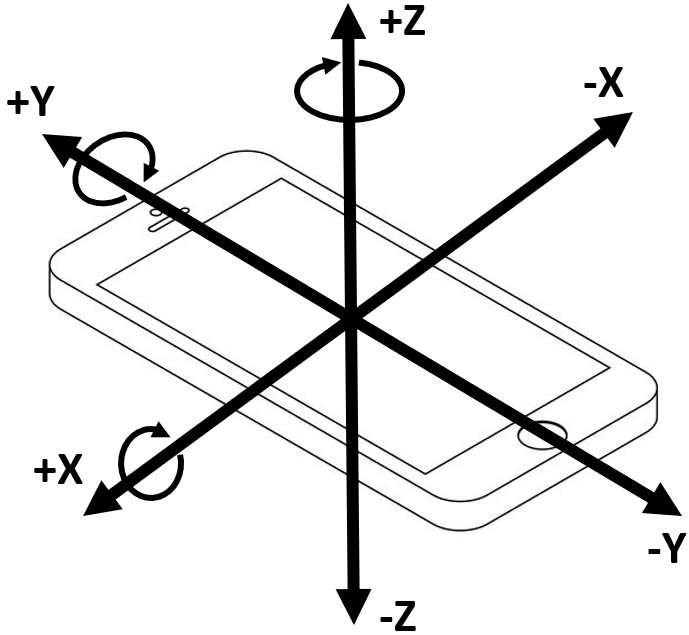}
\caption{The inertial sensor's co-ordinate is aligned with the phone's body frame.}
\label{3dphone}
\end{figure}

Nevertheless, as with other sensor categories, inertial sensors were not designed with indoor positioning in mind. Hence, they face the following challenges.
\begin{itemize}
    \item \textbf{Unconstrained phone orientation}. As the sensor is fixed inside the smartphones, its 3-dimensional coordinates are aligned with the phone's body frame (see Figure~\ref{3dphone}). Hence, we observe different measures with different phone's orientations, in the same spot.
    
    \item \textbf{Uncontrolled body placement.} We have no control on how the user holds their device. As the users move, the placement of the smartphone on their body has a major impact on the sensor measures.
    
    \item \textbf{Unrelated measures}. For positioning purpose, we are only interested in true movement-related force. However, other gestures such as arm, body movements are recorded too.
\end{itemize}

\subsection{Accelerometer}
The accelerometer is designed to measure acceleration forces, which are the changing rate of linear velocity. Such forces may be static such as the Earth's continuous gravitational pull, or dynamic such as the phone vibrations and human movements. The measuring unit is in metre per second squared $(m/s^2)$. The available information is a 3-dimensional vector corresponding to the acceleration force being applied at each Cartesian axis of $(x, y, z)$.

On the smartphone, the accelerometer was originally intended to detect which way the phone is pointing (i.e. based on the Earth's static gravitational pull whose strength and direction are known and constant), in order to adapt the screen into portrait or landscape mode accordingly. Modern smartphones are capable of high sampling rate of up to 500 Hz to detect the slightest change in movements.

\subsection{Gyroscope}
The gyroscope is designed to measure rotary motion, which is the changing rate of the angular rotational velocity. The measuring unit is in degree per second (\textdegree/s). The available information is a 3-dimensional vector corresponding to the rate of rotation around the three Cartesian axes $(x, y, z)$.

On the smartphone, gyroscope was designed to measure how much the phone is titling. On a quick thought, this is what an accelerometer could achieve. However, the critical difference is that an accelerometer may only determine the phone's orientation when it is static, by analysing the Earth's gravitational pull~\cite{nguyen2019realtime}. When the phone is moving, the true acceleration is mixed with the Earth's gravity, which cannot be easily separated. On the other hand, the gyroscope can measure precisely the phone's rotary angle in 3-D. By combining accelerometer and gyroscope, the phone can better sense its position for many motion-sensing applications such as Virtual Reality headset and image stabilisation. Similar to accelerometer, modern smartphones are capable of high gyroscope sampling rate.




\subsection{Positioning methods}
Having introduced the inertial sensors, we will now overview the most popular positioning methods. In principal, there are two main approaches for inertial tracking, that are strap-down, and step-and-heading.

However, the shared challenge for both approaches, as discussed earlier, is that while the sensor measures are aligned to the phone body frame, the phone itself is not fixed to any particular axis, and is likely to be shifted around in the hand or pocket as the user walks. Hence, to work out the true acceleration, we use the gyroscope to track the orientation of the phone, and in doing so, being able to maintain a consistent global reference frame for the sensor measures.

\subsubsection{Tracking the phone orientation}
The steps to track the phone orientation are as follows. Firstly, we need to convert the sensor's local body frame, which changes as the phone is rotated, into a global fixed frame. Given the angular rotational velocity provided by the gyroscope at time point $t$: $g(t) = (g_x(t), g_y(t), g_z(t))$, we specify the following 3x3 rotation matrices $O_x, O_y, O_z$, where each column represents a vector along the global frame axis.
\begin{equation}
    \begin{aligned}
        O_x(t) = 
        \begin{bmatrix}
            1 & 0 & 0 \\
            0 & -cos(g_x(t)) & sin(g_x(t)) \\
            0 & sin(g_x(t)) & cos(g_x(t))
        \end{bmatrix} \\
        O_y(t) = 
        \begin{bmatrix}
            cos(g_y(t)) & 0 & sin(g_y(t)) \\
            0 & 1 & 0 \\
            -sin(g_y(t)) & 0 & cos(g_y(t))
        \end{bmatrix} \\
        O_z(t) = 
        \begin{bmatrix}
            cos(g_z(t)) & sin(g_z(t)) & 0 \\
            -sin(g_z(t)) & cos(g_z(t)) & 0 \\
            0 & 0 & 1
        \end{bmatrix}
    \end{aligned}
\end{equation}

Then, the new sensor frame is computed as $O_T = O_x O_y O_z$. To transform the new sensor measures to the global frame, we multiply them with $O_T$~\cite{correa2017review,woodman2007introduction}.

Once the sensor measures are in the same consistent global frame, we can safely subtract the Earth's gravity vector $g = (0, 0, 9.81)$, which is always pointing down on the z-axis, from the accelerometer vector $a(t) = (a_x(t), a_y(t), a_z(t))$, to reveal the true acceleration of the phone at time point $t$, as follows:
\begin{equation}
    a(t)_{noG} = a(t) - g
\end{equation}

The above true acceleration vector of the phone can then be used with the strap-down and step-and-heading approach below.

\subsubsection{Strap-down approach}
With strap-down approach, the phone positions are derived by continuously estimating the velocity based on the measure provided at fixed periods by the accelerometer. 

Without loss of generality, the velocity $v(t)$ of the moving phone at time point $t$ can then be computed as follows:
\begin{equation}
    v(t) = v(0) + \int_{0}^{t} a(t)_{noG} dt ,
\end{equation}
with $v(0)$ = 0 is the initial velocity when the device is static.

The above velocity is then integrated again to produce the phone's position.
\begin{equation}
    p(t) = p(0) + \int_{0}^{t} v(t) dt ,
\end{equation}
with $p(0)$ = 0 is the initial position.

\subsubsection{Step-and-heading approach}
With step-and-heading (SaH) approach, the phone positions are estimated via the user's continuous walking steps and strides. Therefore, this approach is pedestrian specific. In general, SaH contains 3 steps:

\textbf{Step 1: Extracting step related measures}. As the sensor measures contain the acceleration and rotation of the phone in general, the first step is identifying which segments of the sequence represents the walking steps (i.e. to count how many steps the user has walked). There are two options for this task. 
\begin{itemize}
    \item Stance detection: This approach identifies the sensor segment in which the user foot is planted on the floor (i.e. to be counted as one step). This is done by simply checking if the sensor measure is static (i.e. below a certain threshold) which implies the user is not moving at that moment.
    \item Step cycle detection: This approach looks for repeated patterns within the sensor sequence, assuming that the user has a relatively consistent gait through out.
\end{itemize}

\textbf{Step 2: Estimating step length}. Once the relevant step related sensor sequence has been extracted above, each segment will be counted as one walking step. The length of each step can simply be defined as a constant (e.g. 70 cm). This is a relatively practical assumption for natural walking pace~\cite{harle2013survey}. However, the human stride may vary significantly by up to 50\% when the walking speed changes~\cite{weinberg2002using}.

\textbf{Step 3: Estimating heading}. The user's heading is then inferred in a similar manner as in strap-down system, by integrating the gyroscope measure to obtain the heading change estimation. However, the advantage is that there is no need for a second integration of the accelerometer measure to work out the travelling distance as we may just use the step length above. Therefore, the accumulated error grows linearly, rather than cubically in strap-down approach.

Without loss of generality, the position of the moving phone at time point $t$ can be computed as follows:
\begin{equation}
    p(t) = p(t-1) + step * cos(\theta(t))
\end{equation}
where $step$ is the step length and $\theta$ is the heading.

\begin{figure}[h]
    \centering
    \includegraphics[width=3.2in]{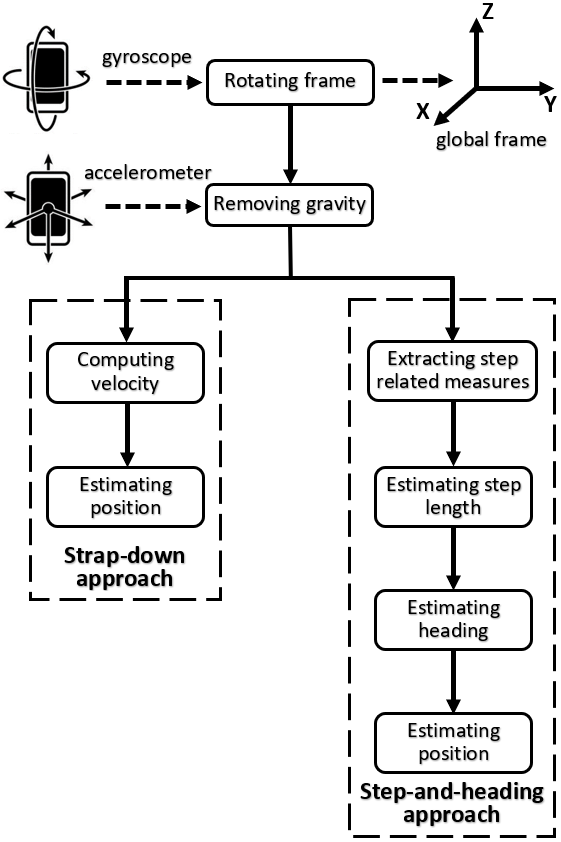}
    \caption{The two main approaches for inertial based system.}
    \label{inertialprocess}
\end{figure}

Figure~\ref{inertialprocess} visually summarises the processes in an inertial tracking system.


\subsubsection{Incorporating floor-plan information}
With inertial tracking systems, the user position is estimated sequentially, in relation to the starting point (whereas wireless signal based or visible light based systems can estimate the user position independently at any moment without reference to previous estimations). Therefore, any error happening during the process (which unfortunately is inevitable at both hardware and software levels, because of the sensor orientation and the integration in the algorithm) will accumulate rapidly, the longer the user trajectory is.

As such, inertial systems need other source of information to correct these position drifts. One such information is the floor-plan of the tracking zone, which details the geometry shape of the walls, rooms, corridors. The working principle is that the transition between two consecutive positioning estimations must not violate the geometric constraints imposed by the floor-plan (i.e. the user cannot walk through walls, or jump to the other side of the corridor). There are two popular implementations of such information in the literature.
\begin{itemize}
    \item \textbf{Refining the user positions.} If the starting point on the map is known before-hand, the system just actively monitors the position estimation at each step, making sure it is within the boundaries on the map.
    
    \item \textbf{Constraining the walking trajectory.} If the starting point is unknown, the system builds up the shape of the trajectory along the way. At some point, the path is long and distinctive enough to be matched against the floor-plan to provide an estimation of the user's whereabouts.
\end{itemize}

\subsection{Performance review}
\label{performanceinertial}
Having introduced the inertial based techniques, we are now in a good position to review their performances on real world systems (see Table~\ref{comparisoninertial}).

Technique wise, strap-down approach performed slightly worse than step-and-heading one~\cite{solin2018inertial,kang2012improved}, as it has to perform the double integration, as explained earlier. Since the phone is not fixed to the user body, any small errors in estimating the acceleration and rotation will translate into progressively bigger error after each integration. Generally speaking, the longer pure inertial based systems are executed, the larger the error will accrue, and the further the estimated position will deviate from the true location, which may be as much as 17 metres after just 550 steps~\cite{faragher2013smartslam}. To tackle this challenge, Zhao et al.~\cite{zhao2019probabilistic} infers the magnetometer data to reduce the heading drift, whereas Solin et al.~\cite{solin2018inertial} exploits the occasional user stationary states to reset the velocity. Nevertheless, the indoor magnetic field may be heavily distorted indoors, and the user may keep walking for long period of time~\cite{nguyen2017dynamic}.


\begin{table*}[!ht]
	\caption{Comparison of notable inertial based system performance.}
	\centering
	\begin{tabular}{C{2cm}C{2.5cm}C{2cm}C{2.5cm}C{2cm}C{5cm}}
		\toprule
		\textbf{Authors} & \textbf{Sensor} & \textbf{Positioning technique} & \textbf{Testbed} & \textbf{Accuracy} & \textbf{Notes} \\
		\midrule
		
		Kang et al.~\cite{kang2012improved} & Accelerometer, Magnetometer, Gyroscope & Step-and-heading & 85m walking path, 1 minute walk & 1.23m mean error & The authors propose to select the most likelihood time between 2 consecutive steps where the heading direction may change. They tested their system on a smaller 75 meter walking path, with slightly better accuracy. \\ \addlinespace[0.2cm]
		
		Zhao et al.~\cite{zhao2019pedestrian} & Accelerometer, Magnetometer, Gyroscope & Step-and-heading & 858m walking path, 664 steps & 1.57m mean error($\pm$0.95m) & Gradient descent algorithm was implemented to reduce the heading drift using the magnetometer data which was distorted by hard-iron. \\ \addlinespace[0.2cm]
		
		Solin et al.~\cite{solin2018inertial} & Accelerometer, Gyroscope & Strap-down & ~40 metre path, 2 minute walk & 1.8m mean error & An extended Kalman filter was applied to fuse the accelerometer and gyroscope measures. The model is constrained by zero-velocity updates when the user is stationary. \\ \addlinespace[0.2cm]
		
		Qian et al.~\cite{qian2013improved} & Accelerometer, Magnetometer, Gyroscope & Step-and-heading & 18m x 12m lab, 60m walking path & 1.71m, 95\% chance, phone in pocket & The floor-plan was used to constrain the inertial drifting. The positioning accuracy was higher at 1m, 95\% chance when the phone was in hand.\\ \addlinespace[0.2cm]
		
	    
		\bottomrule
	\end{tabular}
	
	\label{comparisoninertial}
\end{table*}

\subsection{Overall remarks and open research questions}
Inertial based approach's appeal lies in the scalability as the system is self-contained and requires no additional infrastructure. Availability wise, almost every single smartphone comes equipped with inertial sensors to support the basic functions of adapting the phone screen's orientation. Accessibility wise, from the indoor positioning service's perspective, this group of sensors can be accessed at any given time by the app, without any user permission, to deliver a smooth user experience, yet some may argue about its potential security risk~\cite{nguyen2019location}. Power wise, inertial sensors belong to the group of so-called low power sensors, which consume so little power that Android allows them to be always on.

Positioning accuracy wise, inertial systems achieve around 1 metre on average, which is great considering its advantages above. However, such accuracy quickly degrades the longer the user walks. Hence, without ground-truth data to correct the sensor and estimation error, such systems may not be usable in the long run.

Research wise, potential academic researchers may be interested in the following open questions:
\begin{itemize}
    \item \textbf{How to detect various step lengths?} Most inertial systems assume a constant step length through out the user's journey, which may not be true for different walking paces. As stride length is an important parameter in the positioning estimation equation, it should be adjusted in real time.
    
    \item \textbf{How to robustly estimate the sensor heading?} The user may hold the phone in hand, shirt pocket, trouser pocket, bag, etc. which impacts the sensor measures. The magnetometer which was designed to track the phone's heading, does not work reliably indoors because of the high magnetic anomalies.
    
    \item \textbf{How to address sensor drifting?} This is perhaps one of the biggest challenges. In addition, the phone sensors are noisy which amplifies the estimation errors.
\end{itemize}


\section{Other systems}
This section overviews the sensors that do not fit directly in the above three categories. Unfortunately, at the time of writing, most of these sensors are not explicitly useful for indoor positioning, but we include them here for completeness.

\subsection{Microphone}
The microphone is an acoustic sensor that converts sound waves into electrical signals. In contrast to the previously discussed electromagnetic waves that can propagate seamlessly through space, sound waves are mechanical waves that require a material medium such as air or water to propagate from one place to another.

On the smartphones, the microphone was designed for human speech recording. Hence, its sensitivity was factory-adjusted to specifically aim at human audible voice ranging from 20 Hz to 20 kHz. However, some microphones could pick up inaudible sounds beyond this range~\cite{roy2017backdoor,deshotels2014inaudible}. The available information is the raw sound data, and the maximum amplitude of the noise which can be converted into decibel unit. Modern smartphones may have 2 microphones, one mouthpiece located at the bottom of the phone, where it is closest to the human mouth for the loudest voice recording, and a back microphone located in the back of the phone to capture the ambient sound. The combination of two audio sources will help with noise reduction and improving sound quality.

The clear advantage of microphone is that it is a ubiquitous sensor on every single smartphone to facilitate the essential function of audio communication. Nevertheless, it faces the following challenges.
\begin{itemize}
    \item \textbf{Audio level.} High frequency noise can be easily blocked by covering the microphone (e.g. when it is inside pocket or bag). Similarly, the clarity of the audio recording can be impacted by changing the sensor orientation.
    
    \item \textbf{Audio distortion.} Microphone has an volume level upper-bound which it may handle. When the incoming sound is too loud, the captured recording may be distorted.
    
    \item \textbf{Low temporal variation.} The sound within an indoor space tends to vary significantly over time by infrequent noises (e.g. people chatting occasionally, machine noise during daytime but not nighttime). This has a major impact on systems mapping sound to location (e.g. fingerprinting).
\end{itemize}

Overall, microphone offers a means to record sound, which may be useful for some indoor navigation systems.

\subsection{Fingerprint}
Fingerprint sensor is designed to capture the friction ridges and valleys on the human's fingers. The sensor can either be optical based or capacitive based. Optical sensor works by shining a bright LED light over the fingerprint, and taking an image of it. Capacitive sensor, on the other hand, uses electrical current to capture the fingerprint, which is more robust against LED, and makes the system harder to spoof with fake fingerprints.

At the time of writing, due to the very short working range, which requires the finger to be placed right on top of the sensor, it does not appear to be suitable for indoor positioning.

\subsection{Barometer}
The barometer is designed to measure air pressure (i.e. the weight of the air exerted by the overhead atmosphere onto a surface or object). The sensor was originally meant for weather forecasting, as air pressure is caused by the amount of water within the air. The higher the reading is, the better the weather will be. In contrast, the lower the reading is, the more likely it will rain. The measuring unit is Pascal (Pa). 

On the smartphone, however, barometer is purposely used to improve the GPS accuracy with the added altitude information. The observation is that the atmospheric pressure decreases as altitude increases. There exist many models to map altitude to pressure~\cite{hobbs2000introduction,li2013using}. However, as barometer was not originally designed for such purpose, it faces the following challenges.
\begin{itemize}
    \item \textbf{Closed environment.} Barometer works best in clear, open air. The air indoors circulate in a closed area, where air-conditioner, fans may impact the reading.
    
    \item \textbf{Low sensitivity.} The sensor measure does not vary significantly with respect to small altitude changes. It was reported that the pressure drops just 11 Pa for each metre~\cite{li2013using}. Therefore, this measure may be lost amongst other sensor noises.
\end{itemize}

Overall, barometer is useful to distinguish users on different floors in buildings with high ceiling, rather than determining fine-grained positions on the same floor level. It adds the extra z-dimension into the positioning information to complement other systems.
\begin{table*}[!ht]
	\caption{Comparison of notable sound based system performance.}
	\centering
	\begin{tabular}{C{2cm}C{2.5cm}C{2cm}C{2.5cm}C{2cm}C{5cm}}
		\toprule
		\textbf{Authors} & \textbf{Sensor} & \textbf{Positioning technique} & \textbf{Testbed} & \textbf{Accuracy} & \textbf{Notes} \\
		\midrule
		
		Tarzia et al.~\cite{tarzia2011indoor} & Microphone & Fingerprinting & 43 rooms, 2 training positions per room & 60\% room-level accuracy & Natural ambient noise was recorded, then coverted into power spectrogram to generate a training database. Nearest neighbour using Euclidean distance metric was employed for location matching.\\ \addlinespace[0.2cm]
		
		Rossi et al.~\cite{rossi2013roomsense} & Microphone, Speaker & Fingerprinting & 1 training position every $9m^2$, 67 in total across 20 rooms & 98\% room-level accuracy, 51.3\% within-room accuracy & The smartphone speaker was used to broadcast noise at different locations to generate a training database. SVM was employed for location matching. \\ \addlinespace[0.2cm]
		
		Rishabh et al.~\cite{rishabh2012indoor} & Microphone, Speaker & Trilateration & 6m x 5m x 2.6m room, 6 ceiling mounted speakers & 50cm, 90\% chance & Standard PC speakers were deployed to play in-audible low frequency noises. A sound propagation model was employed to estimate the distance from the smartphone microphone to the speakers. \\ \addlinespace[0.2cm]
		
		\bottomrule
	\end{tabular}
	
	\label{comparisonothers}
\end{table*}

\subsection{Thermometer}
Thermometer is a sensor used to measure the temperature. It is based on the concept of metal resistance (i.e. the flow of electricity changes as the temperature around it changes).

All smartphones do come equipped with this sensor. However, it is included to measure the temperature of the internal components (e.g. battery, CPU chip, etc.) within the phone to protect them from overheating, rather than for the ambient room temperature.

The challenge for having room thermometer on smartphones is that the phone body which houses the sensor is constantly heating up and cooling down, which will impact the true sensor reading of outside temperature. It would not be practical to have this sensor sticking out of the phone body either. As a result, not many smartphones come equipped with room thermometer (e.g. Samsung S4, and Note 3 released 7 years ago are a very few examples). When they do, their usage is rather limited, including strict instructions asking the user to let the phone cooling off without using it for a long period of time on a wooden surface as the human hand may warm the phone up. It was not surprising that Samsung no longer included this type of sensor ever since.

In its current state, thermometer appears to have little use for indoor positioning, but it is included here for completeness.

\subsection{Performance review}
\label{performanceothers}
This section only reviews the performance of microphone in real world systems, as this is the only sensor in this category, capable of powering the entire positioning system alone (see Table~\ref{comparisonothers}). There are two main types microphone based systems. The first type generates a fingerprinting database using passive ambient sound. The second type deploys additional speakers around the building with different noise levels and patterns, then record a fingerprinting database in a similar manner.

Positioning accuracy wise, systems that made use of dedicated speakers achieved much higher accuracy at just 50 centimetre error, compared to those using natural ambient sound at room-level~\cite{rishabh2012indoor,tarzia2011indoor}. This result is not surprising given most indoor spaces have indistinguishable ambient noise. Rossi et al. proposed an interesting approach where the in-built smartphone speaker is used to broadcast a short acoustic noise, which is then captured by the microphone on the same device, hence avoiding the use of additional infrastructure~\cite{rossi2013roomsense}. Their assumption is that the acoustic impulse response is different for each location, to generate a fingerprinting database. Their results demonstrated 98\% room-level accuracy, 51.3\% within-room accuracy, tested in 20 different rooms.

Overall, most sound based positioning systems offer around room-level accuracy, although those making use of dedicated speakers may offer sub-metre accuracy. Nevertheless, there are concerns regarding the sustainability of such accuracy, as indoor sound's temporal variation is high (e.g. an office is more noisy during daytime than nighttime).

\section{Comparison of technologies}
Having reviewed individual sensor system, we may now compare their performances with each other.

There are several metrics to rank different indoor positioning technologies. As the users may prioritise different criteria depending on their applications, we will review most criteria here, and let our readers decide on which one is suitable for them.

In particular, the criteria for assessment will be accuracy, latency, cost, maintenance, and scalability. Security, despite being an important factor, is not part of this review, as there are many external factors that may impact the system. Table~\ref{comparisonperformance} compares the sensor performances using the above metrics.
\begin{table}[h]
	\caption{System performance under different metrics.}
	\centering
	\begin{tabular}{cccccc}
		\toprule
		\textbf{Sensor type} & \textbf{Accuracy} & \textbf{Latency} & \textbf{Cost} & \textbf{Scalability}\\
		\midrule
		Barometer & floor-level & high & low & high \\ \addlinespace[0.2cm]	
		
		Bluetooth & room-level & high & medium & low  \\ \addlinespace[0.2cm]
		
		Camera & high & low & changing & high \\ \addlinespace[0.2cm]
		
		Cellular & town-level & high & low & high \\ \addlinespace[0.2cm]
		
		GPS	& low & high & low & high \\ \addlinespace[0.2cm]
		
		Inertial & changing & low & low & high \\ \addlinespace[0.2cm]
		
		Magnetometer & medium & low & low & low  \\ \addlinespace[0.2cm]
		
		Microphone & low & low & low & low  \\ \addlinespace[0.2cm]
		
		Time-of-flight & high & high & low & low \\ \addlinespace[0.2cm]
		
		WiFi & medium & high & low & low \\ \addlinespace[0.2cm]
		\bottomrule
	\end{tabular}
	
	* Inertial sensors include the accelerometer and gyroscope.
	\label{comparisonperformance}
\end{table}

\subsection{Accuracy}
Accuracy is arguably the most considered factor for a positioning system. This metric will decide if the system is capable of achieving fine-grained tracking with precise positions, or just coarse-grained proximity tracking in a wide area. It reports the difference between the estimated position by the system and the true position. Some systems average these results over multiple testing locations, others report the variation of the above positioning accuracy (e.g. 2 metre accuracy, 95\% of the time). Normally, the Cumulative Distribution Function (CDF) is employed to report the system's accuracy.

The results in Sections~\ref{performanceelectromagnet},~\ref{performancelight},~\ref{performanceinertial} and~\ref{performanceothers} suggested that visible light based systems achieved some of the highest positioning accuracy at centimetre level, using custom LEDs. This is followed by WiFi, Bluetooth, and Camera systems (using indoor images) at around 2 to 3 metre accuracy. Inertial systems start off rather accurately at sub-metre level, but quickly degrade exponentially the longer the travelling trajectory is. Other systems such as sound based, magnetometer based, and FM based, typically achieved room-level accuracy.

\subsection{Latency}
Latency measures the delay for which the information is provided. This metric is critical for tracking fast moving users. It is a combination of the hardware latency (i.e. how frequent the sensor produces a measure), and the software latency (i.e. the algorithm executing time).

Sensor level wise, inertial sensors have the highest sampling rates at more than 100 Hz (100 samples per second) on most modern smartphones (see Table~\ref{comparisonsensors}). This design was intended to capture fast motions. At the other end of the spectrum were WiFi and Bluetooth, with 1.5 Hz and 0.25 Hz respectively. 

Algorithm level wise, systems that rely on a database may impact the real time positioning speed, if the algorithm requires a full scan of the entire database to estimate the user position (e.g. K-nearest neighbours). A model based approach (e.g. SVM) would accelerate systems involving a large training database. 

\subsection{Cost}
Cost is the total expense to deploy and maintain the positioning system. In general, inertial based approach offers the most cost effective solution, as it needs no additional hardware or software maintenance. The entire system may run locally on the user's smartphone. Systems that make use of ubiquitous infrastructure (e.g. WiFi, magnetic field, etc.) have no infrastructure cost, but may require intensive surveying effort to generate the training database, not least when the tracking zone is large, as well as the maintenance overhead to update such database over time.

\begin{figure}[h]
    \centering
    \includegraphics[width=3.3in]{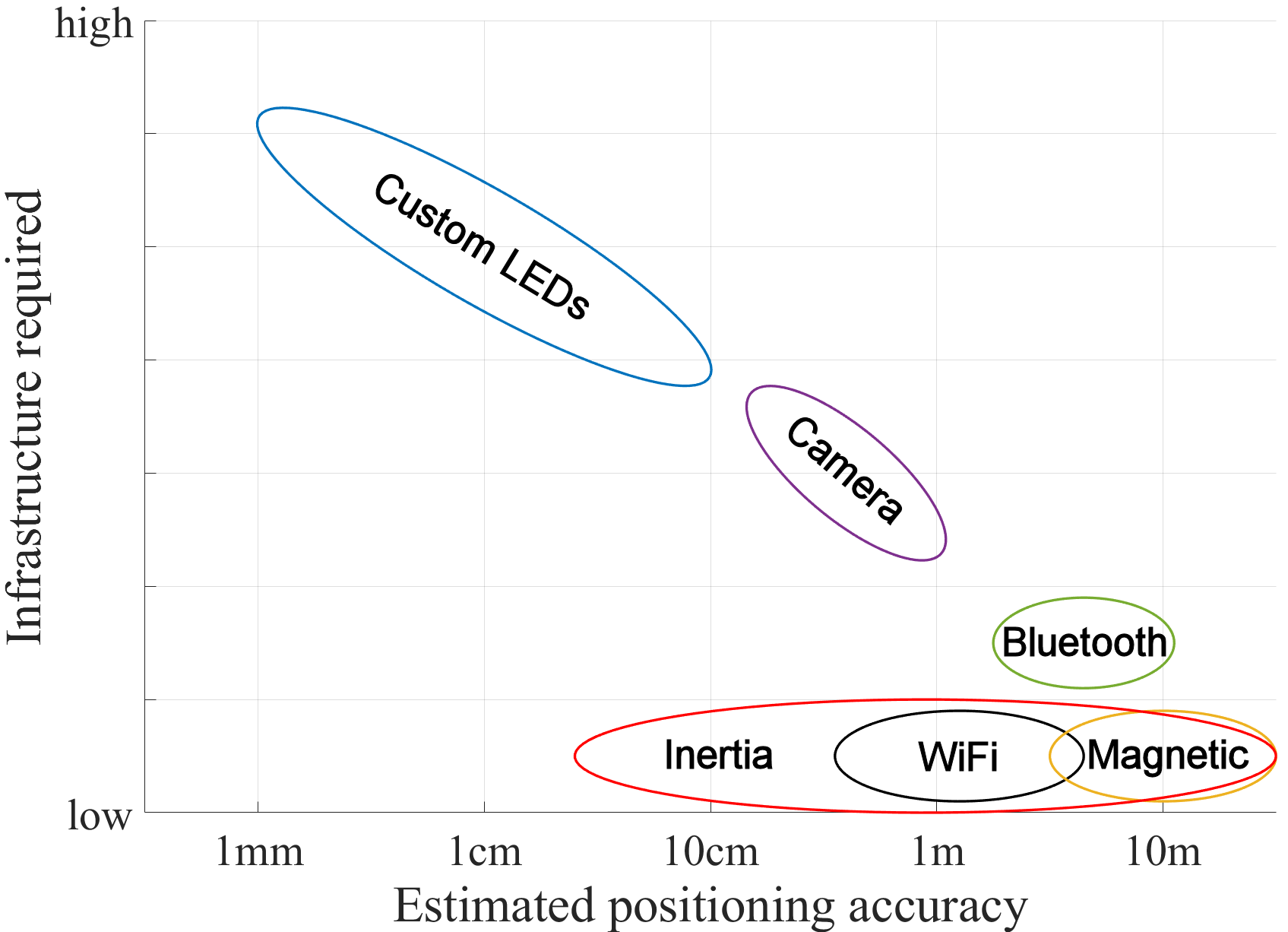}
    \caption{A rough correlation between the amount of infrastructure and the estimated accuracy of indoor positioning technologies. The x-axis is in log scale.}
    \label{infrastructurevscost}
\end{figure}

The general trade-off is that the higher the system cost, the more accurate the system may be, and vice-versa (see Figure~\ref{infrastructurevscost}).

    
\subsection{Scalability} 
This metric is the system's capability to serve a huge number of users simultaneously, with plenty of room for future expansion. This is the most essential metric for commercial solutions, but is perhaps one of the most under-considered metrics for indoor positioning systems.

Generally speaking, scalable systems should function in a decentralised fashion, where the user's smartphones perform most of the work. For example, systems that operate locally on the smartphones (e.g. interial based, custom LEDs) may easily scaled to the vast number of users. On the other hand, systems using a central server (e.g. fingerprinting based) may struggle to serve many users simultaneously, including the bandwidth cost for communication between the users and the server~\cite{nguyen2014selective}.




\section{Hybrid systems}
Having discussed the strengths and weaknesses of individual sensor in previous sections, the overall consensus is that there is no stand-alone best option for all indoor scenarios. As each sensor excels in certain condition, it is not surprising that most effort were invested to combine such sensors together. This section reviews hybrid systems that involve two or more of the sensors described above, and the fusion techniques behind.
\begin{figure*}[!ht]
\centering
\includegraphics[width=5.8in]{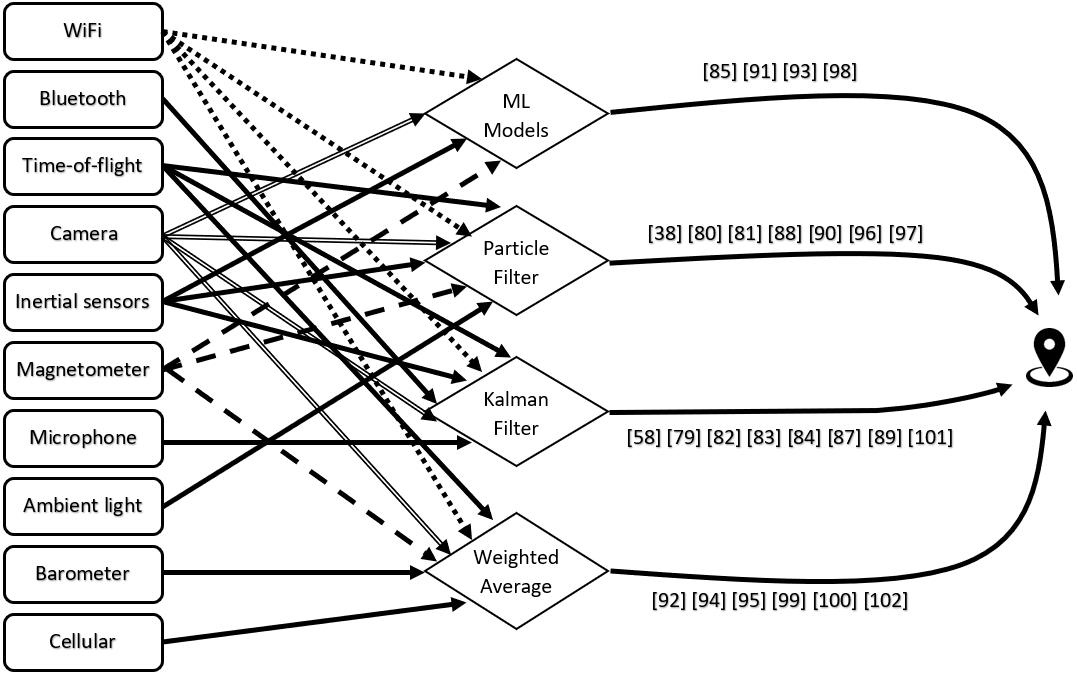}
\caption{Combination of smartphones sensors with respect to the data fusion methods. Different styles of arrows are used to distinguish amongst sensor types and how they combine.}
\label{sensorcombination}
\end{figure*}

\subsection{Sensor fusion techniques}
In indoor positioning, sensor fusion is the process of combining data from several sensors to improve the positioning accuracy. There are three different category in sensor fusion~\cite{king2017application,beddar2020multi}.
\begin{itemize}
    \item \textbf{Complementary category.} With this category, each sensor is responsible for a separate, disjunctive part of the environment, and their measures will later be combined to provide a fuller knowledge of the indoor space. For example, two cameras were set up to cover different indoor areas.
    
    \item \textbf{Co-operative category.} This category uses the measures provided by different sensors to derive new information that was not available before. For example, the combination of camera and time-of-flight sensor provides a new 3-dimensional view of the indoor space.
    
    \item \textbf{Competitive category.} With this category, different sensors of the same type independently provide measures about the same property. For example, two cameras provide two images of the same object. However, one image may be blur due to poorer lens. Hence, the fusion of both images will result in a clearer result with less noise.
\end{itemize}

It is worth emphasising that it is possible to employ more than one approach in the same system, for example, a positioning system using a network of Bluetooth and WiFi beacons deployed at different locations in the building. Each beacon covers a distinct area in a complementary fashion, whereas WiFi and Bluetooth beacons in the same location provide similar distance measure RSS to the smartphone in a competitive fashion. Overall, most smartphones based systems would fall into the co-operative category, because all sensors are housed together within the device. They mostly observe different information of the same indoor space.

Once the sensors' approach has been decided, there are two settings to process their measures as follows.
\begin{itemize}
    \item \textbf{Centralised setting.} In this setting, the measures from all sensors must be available simultaneously during the fusion process.
    
    \item \textbf{Decentralised setting.} In this setting, a model is derived for each sensor. When the fusion happens, each sensor model is processed sequentially.
\end{itemize}

With the smartphones, all sensors can be queried at any given moment. Although they have different sampling rates, faster sensor can be adjusted to match the sampling rate of slower ones. Hence, both of the above settings can be employed for data fusion.

Once the approach and setting have been decided, the fusion algorithms will be applied onto the sensor measures. Within the indoor positioning literature, three popular data fusion algorithms are filtering based (e.g. Particle Filter, Kalman Filter, Simultaneous Localisation and Mapping, etc.)~\cite{faragher2013smartslam,mur2017visual} and machine learning (ML) based (e.g. Hidden Markov Model, Gaussian Mixture Model, Bayesian inference etc.)~\cite{wen2013assessing}, and weighted average. We briefly discuss their core concepts below.
\begin{itemize}
    \item \textbf{Filtering based algorithm.} Filtering algorithms are those that recursively refine a set of positioning estimations over time, given a sequence of sensors' measure. The Kalman Filter, for example, applies linear projections, whereas Particle Filter uses sequential Monte Carlo method~\cite{rhudy2017kalman,faragher2012understanding,an2013prognostics}. Each state normally includes the positioning label, the weight representing the probability of it being the true position, and other set of information related to the sensors being used.
    
    \item \textbf{Machine learning based model.} ML models may be used to represent the distribution of each sensor's measure, which can later be fused using a mixture model (e.g. Gaussian Mixture Model)~\cite{gershman2012tutorial,reynolds2009gaussian}. This is mostly used in the decentralised sensor setting described earlier.
    
    \item \textbf{Weighted average.} With this approach, each sensor produces its own positioning estimation independently. The fusion simply takes the average of the each sensor's results with higher weights awarded to more reliable sensors.
\end{itemize}

Figure~\ref{sensorcombination} visualises the combination of smartphones sensors with respect to different data fusion methods in this review. Two filtering based algorithms, Particle Filter and the Kalman Filter, are separated in the figure for better highlights.

\subsection{Performance review}
With 18 smartphones sensors reviewed above, even after leaving out non-relevant ones to indoor positioning, there are still too many possible permutations of hybrid systems to review. Hence, we mostly focus on combinations with the potential to improve the overall accuracy, where sensors complement for the weaknesses of each other. In particular, some potential matchings at sensor category level are:
\begin{itemize}
    \item \textbf{Inertial and electromagnetic sensors.} Inertial based systems start off very accurately (i.e. within the first few steps), but cannot sustain such performance over long distance as sensor drifting happens. WiFi or Bluetooth beacons, on the other hand, could be used to correct such drifts~\cite{li2015bluetooth,putta2015smartphone,zhu2015novel,maghdid2019indoor}.
    
    \item \textbf{Inertial and visible light sensors.} Similarly, inertial based systems' drifting may be corrected by visual information provided through the camera~\cite{al2013indoor}.
    
    \item \textbf{Visible light and electromagnetic sensors.} VL based systems require sufficient natural light, and good camera viewing angle to function. In contrast, electromagnetic based systems use invisible waves of signal that can function well regardless of the lighting condition~\cite{papaioannou2014fusion,wang2018deepml}.
\end{itemize}

At individual sensor level within each category, some potential matchings are:
\begin{itemize}
    \item \textbf{WiFi and magnetometer.} Distant indoor positions mostly observe different WiFi or Bluetooth fingerprints (as the wireless signals propagate and attenuate), but similar magnetic fingerprints (because of its highly spatial similarity). In contrast, neighbourhood positions tend to have similar WiFi or Bluetooth fingerprints, but different magnetic ones~\cite{nguyen2017co,nguyen2017assessing}. Therefore, they complement each other perfectly.
    
    \item \textbf{Camera and time-of-flight.} Camera captures a broad overview of the surroundings including colour, objects, etc. whereas ToF adds the depth dimension to the image~\cite{lynen2015get}.
\end{itemize}

Tables~\ref{comparisonhybrid1}~and~\ref{comparisonhybrid2} summarise some notable hybrid indoor positioning systems in the recent literature.

\section{Conclusion}
This review has extensively summarised and compared most interesting smartphones based indoor positioning systems and their powered technologies. In doing so, we have listed the most novel open questions for academic researchers, and overviewed some of the most practical systems with mass adoption potentials.

This last section will present some of the authors' personal belief in the current state of indoor positioning, and what the future may hold.

\subsection{How have we been doing for the past 20 years ?}
More and more research articles about indoor positioning are being published yearly, with one claiming to improve on another in terms of accuracy, reliability, practicality, etc. However, after 20 years of research (since some of the very first papers on indoor positioning were published), the most self-reflecting question for every researcher in this domain is: \emph{``Why there is still no prevalent, ready-to-use indoor navigation system in the market yet ?''} The authors believe there are two main factors.
\begin{itemize}
    \item \textbf{Highly accurate and affordable.} Most centimetre level accuracy systems are costly, whereas affordable systems may only manage room-level accuracy. Unfortunately, both of these criteria do not yet co-exist on a single system.
    
    \item \textbf{Changing the user routine.} Most users are not willing to carry an extra piece of hardware, or having to install tracking beacons, or doing maintenance work, just to access the positioning service. Unfortunately, most current indoor positioning systems need one or more of these requirements.
\end{itemize}

The authors took notice that the majority of systems in the literature often prioritise positioning accuracy as the top target for their systems. Although this is no doubt an important criterion, there are other observations.
\begin{itemize}
    \item \textbf{Sensor boundary.} The hardware implementation in all sensors have reached their limitation (i.e. it is nearly impossible to push pure WiFi positioning system based on RSS to consistently achieve centimetre accuracy level, in every building).
    
    \item \textbf{Algorithm efficiency.} The positioning accuracy cannot be improved endlessly with new algorithms, given the same dataset.
\end{itemize}

In short, it is unrealistic to keep exhausting the same set of sensors and algorithms without any fundamental change to the problem in question, and expects the technology to suddenly become widely-adopted one day.

This leads us to the \emph{``what next ?''} question for future indoor positioning researchers, to be addressed in the next section.

\subsection{The future of indoor positioning}
The ultimate research question is: \emph{``What will the future de-facto standard indoor positioning system look like ?''}. Such system may take some hints from the current influential GPS, which are: 
\begin{itemize}
    \item \textbf{Ease of accessibility.} The satellite signals are ready to be accessed virtually everywhere outdoors, without any restriction.

    \item \textbf{Simple calculation.} The user position is computed directly by straight-forward triangulation of the satellite signals.

    \item \textbf{High scalability.} The system design is decentralised, so that the clients work out their position.
\end{itemize}

The futuristic equivalent indoor GPS-like would no doubt share the above features. However, there are two fundamental differences between outdoor positioning and the indoor version. Firstly, GPS was designed to guide the user to the right building (e.g. shop, house) with a high tolerance of 3 to 5 metre error, whereas indoor navigation systems, at minimum, must be able to guide the user to the right room with much lower tolerance of 1 to 2 metre error. Secondly, GPS assumes unobstructed line of view from the satellite to the user which is acceptable outdoors, while the complex indoor interiors make the wireless signal propagate in unexpected fashions.

As such, it may require a complete different approach or new technology for the ideal indoor positioning system. Nevertheless, technologies have evolved rapidly for the past 7 to 8 years, which fortunately has also led to the changes in the user's behaviour.
\begin{itemize}
    \item \textbf{Smartphone as a new accessory}. People carry their smartphones everywhere outdoors, but also indoors for daily routine (e.g. email checking, internet surfing, friend chatting). In addition, many people wear smartwatches (a small form factor of the smartphones) while sleeping for fitness tracking. We have now embraced the smartphones as part of our inseparable necessities.
    
    \item \textbf{Smart buildings.} The indoor infrastructure are improving significantly with modern technologies. Almost every new building is well populated with WiFi APs for wireless communication, and Camera with internet access.
\end{itemize}

In closing, the ubiquity of smartphones and the improving building infrastructure in recent years have presented researchers with new opportunities to develop the ideal future indoor positioning systems, which are not only capable of fine-grained sub-metre accuracy, but are also affordable and simple to operate.

\begin{landscape}
\begin{table}
	\caption{Comparison of notable hybrid system performance.}
	\centering
	\begin{footnotesize} 
	\begin{tabular}{C{1.8cm}C{2.4cm}C{2.2cm}C{3cm}C{2.2cm}C{11.5cm}}
		\toprule
		\textbf{Authors} & \textbf{Sensor} & \textbf{Fusion technique} & \textbf{Testbed} & \textbf{Accuracy} & \textbf{Notes} \\
		\midrule
		
		Zhu et al.~\cite{zhu2015novel} & Inertial sensors, WiFi & Particle Filter & 362.6$m^2$, 4 WiFi APs, 30 training locations & 1.5m, 90\% chance & WiFi fingerprinting and the building map were employed to correct the drifting error from inertial sensors. The proposed method achieved 1.2m mean error, compared to 2.1m using inertial sensors and 2.8m using WiFi.\\ \addlinespace[0.2cm]
		
		Maghdid et al.~\cite{maghdid2019indoor} & Inertial sensors, WiFi & Extended Kalman Filter & 48m x 38m corridor, 7 WiFi APs & 4m, 90\% chance & Inertial tracking alone over 172m travelling distance drifts to over 40m positioning error.\\ \addlinespace[0.2cm]
		
		Li et al.~\cite{li2015bluetooth} & Inertial sensors, Bluetooth & Extended Kalman Filter & 20 Bluetooth beacons over a university corridor & 2.26m mean error & Bluetooth beacons were used for proximity tracking without a training database to correct the inertial sensors' drifting. The positioning results with only 15 and 10 beacons were degraded by 28.9\% and 43.8\% respectively.\\ \addlinespace[0.2cm]
		
		
		
		
		Putta et al.~\cite{putta2015smartphone} & Inertial sensors, Magnetometer & Particle Filter & 9m x 16 m area, 3 walking trajectories of 85 m each & 1.5m, 90\% chance & Gradient descent algorithm was employed to correct the inaccuracies in the user heading estimates due to magnetic perturbations.\\ \addlinespace[0.2cm]
		
		
		Delail et al.~\cite{al2013indoor} & Inertial sensors, Camera & Extended Kalman Filter & 309.4m walking trajectory & 1m mean error & Principal Component Analysis was employed to identify the direction of motion.\\ \addlinespace[0.2cm]
		
		Xu et al.~\cite{xu2015idyll} & Inertial sensors, Ambient light & Particle Filter & 3 buildings with 120 luminaries & 50cm, 90\% chance & Illumination peak detection algorithm was used to distinguish various light intensity amongst locations. The information are then fused with inertial sensors measures.\\ \addlinespace[0.2cm]
		
		Yang et al.~\cite{yang2016smartphone} & Inertial sensors, Microphone & Kalman Filter & 8m x 7m room, 6 ceiling mounted sound receivers & 50cm mean error & Acoustic pulses emitted from the smartphone every 0.3s are captured by the sound receivers to determine the position.\\ \addlinespace[0.2cm]
		
		Shu et al.~\cite{shu2015magicol} & WiFi, Magnetometer & Particle Filter & 47m x 85m office space & 90cm, 90\% chance & The magnetic and WiFi fingerprinting databases were combined under a particle filter for each scan. The result with only the magnetic field was 1.5m, 90\% chance, and 2.5m, 90\% chance with only the WiFi database.\\ \addlinespace[0.2cm]
		
		Shao et al.~\cite{shao2018indoor} & WiFi, Magnetometer & Convolutional Neural Network & 60m x 40m office space & 1m, 90\% chance & The WiFi and magnetic signals were converted into a fingerprint image to represent the features of each location. These images are then input into a deep CNN to predict real-time location estimates.\\ \addlinespace[0.2cm]
		
		Baniukevic et al.~\cite{baniukevic2013hybrid} & WiFi, Bluetooth & Weighted Average & 50$m^2$, more than 30 WiFi APs, 12 training points & 1.75m mean error & Pure WiFi positioning achieved 3.15m mean error. Pure Bluetooth positioning achieved 7.57m mean error. The positioning result in a large shopping centre with less stable signals was about 7m mean error.\\ \addlinespace[0.2cm]
		
		Liu et al.~\cite{liu2012push} & WiFi, Microphone & Graph mapping & 12m x 11m office, 71 training locations, 14 WiFi APs & 1.2m, 90\% chance & When a phone needs to improve its positioning accuracy, it sends a special audio signal to recruit other nearby phones. Acoustic ranging estimates amongst phones were used alongside the WiFi fingerprint map.\\ \addlinespace[0.2cm]
		
		Bisio et al.~\cite{bisio2018wifi} & WiFi, Barometer & Weighted Average & 4 floors, 150$m^2$ \& 28 WiFi APs per floor & 1.5m mean error & Barometer measure was much more effective when fewer APs were detected. WiFi positioning only achieved 4.1m mean error. Adding barometer would increase the result to 2.5m mean error.\\ \addlinespace[0.2cm]
		
		Xu et al.~\cite{xu2015enhancing} & WiFi, Camera & Weighted Average & shopping mall, 1.8m space in training points & 2.2m, 90\% chance & Geometric constraints were extracted from images to reduce WiFi fingerprint ambiguity by mapping the constraints jointly against the fingerprint space.\\ \addlinespace[0.2cm]
		
		\bottomrule
	\end{tabular}
	\end{footnotesize}
	
	\label{comparisonhybrid1}
\end{table}
\end{landscape}

\begin{landscape}
\begin{table}
	\caption{Comparison of notable hybrid system performance.}
	\centering
	\begin{footnotesize} 
	\begin{tabular}{C{1.8cm}C{2.5cm}C{1.9cm}C{3cm}C{2.3cm}C{11cm}}
		\toprule
		\textbf{Authors} & \textbf{Sensor} & \textbf{Fusion technique} & \textbf{Testbed} & \textbf{Accuracy} & \textbf{Notes} \\
		\midrule
		
		Papaioannou et al.~\cite{papaioannou2014fusion} & WiFi, Camera & Kalman Filter & 11m x 12m area, 1 camera mounted 10m off the ground & 1m, 90\% chance & The user trajectory is estimated through a series of WiFi scans, fused with the images provided by the ceiling mounted camera. At 60 frame window, the positioning accuracy using both WiFi and Camera was 3 times higher than that with only the Camera information.\\ \addlinespace[0.2cm]
		
		Wang et al.~\cite{wang2018deepml} & Camera, Magnetometer & Long-Short Term Memory & 12m x 6m lab & 3.5m, 90\% chance & Bimodal images were built to train the deep LSTM network. The magnetic field and light data were then used to estimate the user location.\\ \addlinespace[0.2cm]
		
		Liu et al.~\cite{liu2016fusion} & Camera, Magnetometer & Particle Filter & 4,094$m^2$ lab & 1.34m, 90\% chance & A fingerprinting database of the magnetic field and image of each location was generated off-line. Convolutional neural network was employed to extract deep features from the images, which are then fused with the magnetic signals.\\ \addlinespace[0.2cm]
		
		Zou et al.~\cite{zou2017accurate} & Inertial sensors, WiFi, Bluetooth & Particle Filter & 600$m^2$ office, 8 Bluetooth beacons, 8 WiFi APs & 1.1m, 90\% chance & Under the same testbed, WiFi alone achieved 3.1m, 90\% chance, whereas inertial tracking quickly degraded beyond 4m after just 168 steps.\\ \addlinespace[0.2cm]
		
		Faragher et al.~\cite{faragher2013smartslam} & Inertial sensors, WiFi, Magnetometer & Extended Kalman Filter & 45m long office corridor & 2.7m, 95\% chance & WiFi and magnetic fingerprinting were used to constrain inertial sensors' drifting. The positioning error with only the inertial sensors degraded to 16m after 520 steps.\\ \addlinespace[0.2cm]
		
		Ban et al.~\cite{ban2015indoor} & Inertial sensors, WiFi, Magnetometer & Gaussian Mixture Models & 118 m walking trajectory in a station & 6m mean error, 77\% floor accuracy & The WiFi and magnetic fingerprints were represented by a GMM. The positioning accuracy was slightly lower at 7.6m mean error, 74\% floor accuracy, when the phone was in pocket. \\ \addlinespace[0.2cm]
		
		Lynen et al.~\cite{lynen2015get} & Inertial sensors, Camera, Time-of-flight & Extended Kalman Filter & 1.44 km walking trajectory, 703,362 3D points & 25cm mean error & The 3D point cloud is generated off-line from a set of database images. \\ \addlinespace[0.2cm]
		
		Gu et al.~\cite{gu2017waipo} & WiFi, Camera, Magnetometer & Weighted Average & 90m x 50m lab, 25 rooms & 1m, 90\% chance & The users' spatio-temporal co-occurrence information were combined with the WiFi and magnetic fingerprints. Bluetooth was employed to detect users' co-location. Position information are combined for users staying in the same room.\\ \addlinespace[0.2cm]
		
		Kim et al.~\cite{kim2014multi} & WiFi, Magnetometer, Cellular & Weighted Average & 2,000$m^2$ lab, 3x3$m^2$ grid for training & 3.5m, 90\% chance & A fingerprinting data of the WiFi, magnetic, and cellular signals was generated. The search space was sequentially reduced by first applying WiFi fingerprints, then cellular fingerprints, and finally the magnetic fingerprints.\\ \addlinespace[0.2cm]
		
		Ruotsalainen et al.~\cite{ruotsalainen2011visual} & WiFi, Bluetooth, GPS, Inertial sensors, Camera & Extended Kalman Filter & office building & 3.4m mean accuracy & The positioning accuracy in the same office, with occasional GPS data was 13.1m, with Bluetooth was 6.3m, with WiFi was 8.9m, mean error.\\ \addlinespace[0.2cm]
		
		Nguyen et al.~\cite{nguyen2017assessing} & Camera, Time-of-flight, WiFi, Magnetometer & Particle Filter & 900$m^2$ auditorium & 3.5m, 90\% chance & The positioning error in a separate office corridor was only 50cm, 90\% chance. Using only WiFi, the result was 2.5m, 90\% chance.\\ \addlinespace[0.2cm]
		
		Du et al.~\cite{du2016camera} & Camera, Magnetometer, Time-of-flight & Weighted Average & 335$m^2$ lab & 1.6m mean error & The positioning accuracy in the same office, with only the magnetic field 3.1m, with the Camera was 2.1m, mean error.\\ \addlinespace[0.2cm]
		
		
		\bottomrule
	\end{tabular}
	\end{footnotesize}
	
	\label{comparisonhybrid2}
\end{table}
\end{landscape}


\vskip2pc



\bibliographystyle{iet}
\bibliography{references}
%


\end{document}